\documentclass[10pt,journal,compsoc]{IEEEtran}
\usepackage[ruled,linesnumbered]{algorithm2e}
\usepackage{textcomp}
\usepackage{graphicx}
\usepackage{balance}
\usepackage{amsmath}
\usepackage{dsfont}
\usepackage{amsthm}
\usepackage[hyphens]{url}
\usepackage{breakurl}
\usepackage[justification=centering]{caption}
\usepackage{amssymb}
\usepackage[labelformat=simple]{subcaption}
\usepackage{color} 
\ifCLASSOPTIONcompsoc
\usepackage[nocompress]{cite}
\else
\usepackage{cite}
\fi
\ifCLASSINFOpdf
\else
\fi
\begin{document}
	%
	\title{Optimizing Mobile-Friendly Viewport Prediction for Live 360-Degree Video Streaming}
	%
	%
	%
	%
	
	\author{Lei~Zhang,~\IEEEmembership{Member,~IEEE,}
                Tao~Long,
			Weizhen~Xu,
		Laizhong~Cui,~\IEEEmembership{Senior~Member,~IEEE,}
		and~Jiangchuan~Liu,~\IEEEmembership{Fellow,~IEEE}
		
		\IEEEcompsocitemizethanks{\IEEEcompsocthanksitem L. Zhang, T. Long, W. Xu, and L. Cui are with the
			College of Computer Science and Software Engineering, Shenzhen University, Shenzhen, Guangdong 518060, China. 
			E-mail: leizhang@szu.edu.cn;longtao2021@email.szu.edu.cn; xuweizhen2020@email.szu.edu.cn; cuilz@szu.edu.cn.
			\IEEEcompsocthanksitem J. Liu is with the School of Computing Science, Simon Fraser University, Burnaby, BC V5A 1S6, Canada. 
			E-mail: jcliu@cs.sfu.ca.
		}
		
	}
	
	%
	%

\markboth{IEEE TRANSACTIONS ON MOBILE COMPUTING,~Vol.~XX, No.~XX, XX~XX}%
{Shell \MakeLowercase{\textit{et al.}}: Bare Demo of IEEEtran.cls for Computer Society Journals}
%



\IEEEtitleabstractindextext{%
	\begin{abstract}
		Viewport prediction is the crucial task for adaptive 360-degree video streaming, as the bitrate control algorithms usually require the knowledge of the user's viewing portions of the frames. Various methods are studied and adopted for viewport prediction from less accurate statistic tools to highly calibrated deep neural networks. Conventionally, it is difficult to implement sophisticated deep learning methods on mobile devices, which have limited computation capability. In this work, we propose an advanced learning-based viewport prediction approach and carefully design it to introduce minimal transmission and computation overhead for mobile terminals. 
		We also propose a model-agnostic meta-learning (MAML) based saliency prediction network trainer, which provides a few-sample fast training solution to obtain the prediction model by utilizing the information from the past models. 
		We further discuss how to integrate this mobile-friendly viewport prediction (MFVP) approach into a typical 360-degree video live streaming system by formulating and solving the bitrate adaptation problem. Extensive experiment results show that our prediction approach can work in real-time for live video streaming and can achieve higher accuracies compared to other existing prediction methods on mobile end, which, together with our bitrate adaptation algorithm, significantly improves the streaming QoE from various aspects.\textcolor{black}{ We observe the accuracy of MFVP is 8.1$\%$ to 28.7$\%$ higher than other algorithms and achieves 3.73$\%$ to 14.96$\%$ higher average quality level and 49.6$\%$ to 74.97$\%$ less quality level change than other algorithms. }
	\end{abstract}
	
	}

\maketitle

\IEEEdisplaynontitleabstractindextext

%
\IEEEpeerreviewmaketitle

\IEEEraisesectionheading{\section{Introduction}\label{sec:introduction}}

%
%
%
%

\IEEEPARstart{W}{ith}
the rapid growth of personal multimedia devices and the advance of immersive multimedia technology such as virtual reality (VR) and augmented reality (AR), 360-degree videos are becoming more popular than ever before. 
Unlike conventional videos, 360-degree videos can provide panoramic views but cost much more bandwidth for streaming~\cite{qian2018flare}. 
As users have the flexibility to choose which part of the 360-degree scene to watch (referred to as viewport/Field-of-View), numerous existing studies investigate the viewport adaptive streaming approach to improve bandwidth efficiency and enhance user experience. Tile-based approach~\cite{xie2017360probdash} is widely adopted for 360-degree video streaming, which divides each panoramic frame into smaller-sized non-overlapping rectangular regions called tiles. As each tile is independently decodable, the clients can only request or assign higher bitrates to the tiles that are predicted to be in the user viewport. In general, tile-based streaming exploits the trade off between bandwidth efficiency and user experience.

There is no doubt that the most important and the most challenging task for the tile-based 360-degree video adaptive streaming is viewport prediction, which directly affects the tile selections for download and the bitrate decisions from the rate adaptation algorithms. When the viewports are accurately predicted, we can fully allocate the bandwidth resources to the tiles with the video contents that are watched by the user. Unfortunately, prediction errors always exist. If a tile of non-sight is predicted in the viewport, downloading it does not improve the user's quality of experience (QoE), which thus wastes bandwidth. It is even worse when a tile in the viewport is not included in the prediction result, which will cause severe video quality degradation or even a playback interruption for rebuffering the missing tile.

Since predicting human behavior is difficult by nature, viewport prediction has attracted considerable research interests. To design a viewport prediction scheme, a proper machine learning (ML) technique is usually chosen as the basis of the prediction. In general, there are two types of choices: (1) light but probably less accurate algorithms (e.g., regression-based) and (2)  complicated but more accurate algorithms (e.g., deep neural network-based). The algorithms of the first type are adopted in most real-time streaming systems for mobile clients, which can run fast enough under short prediction windows with less overhead. The algorithms of the second type usually introduces substantial computation demands and significant training time, which implies huge implementation overhead and makes it less practical for mobile end devices. Therefore, on one hand, the lightweight viewport prediction algorithms are widely used but less accurate; on the other hand, the sophisticated deep learning-based algorithms can achieve better prediction performance but hard to be realized on mobile clients for 360-degree video live streaming.

In this work, we aim at better bridging the gap between viewport prediction performance and its feasibility on mobile devices. To fulfill our goal, we are facing three challenges. First, as the most important task is viewport prediction, we need to have a comprehensive learning framework that can utilize as much information about video watching as possible to make highly accurate predictions. Second, the training and inference tasks for learning should be carefully distributed, otherwise the deep learning-based approach can cause excessive costs on mobile devices (e.g., long inferring time that is unacceptable in live video streaming). Third, to enable such a learning-based viewport prediction, transmission overhead and computation overhead are inevitably introduced to the 360-degree video streaming system, which should be carefully considered and minimized for practical usage.

\begin{figure*}
	\centering
	\includegraphics[width=0.85\linewidth]{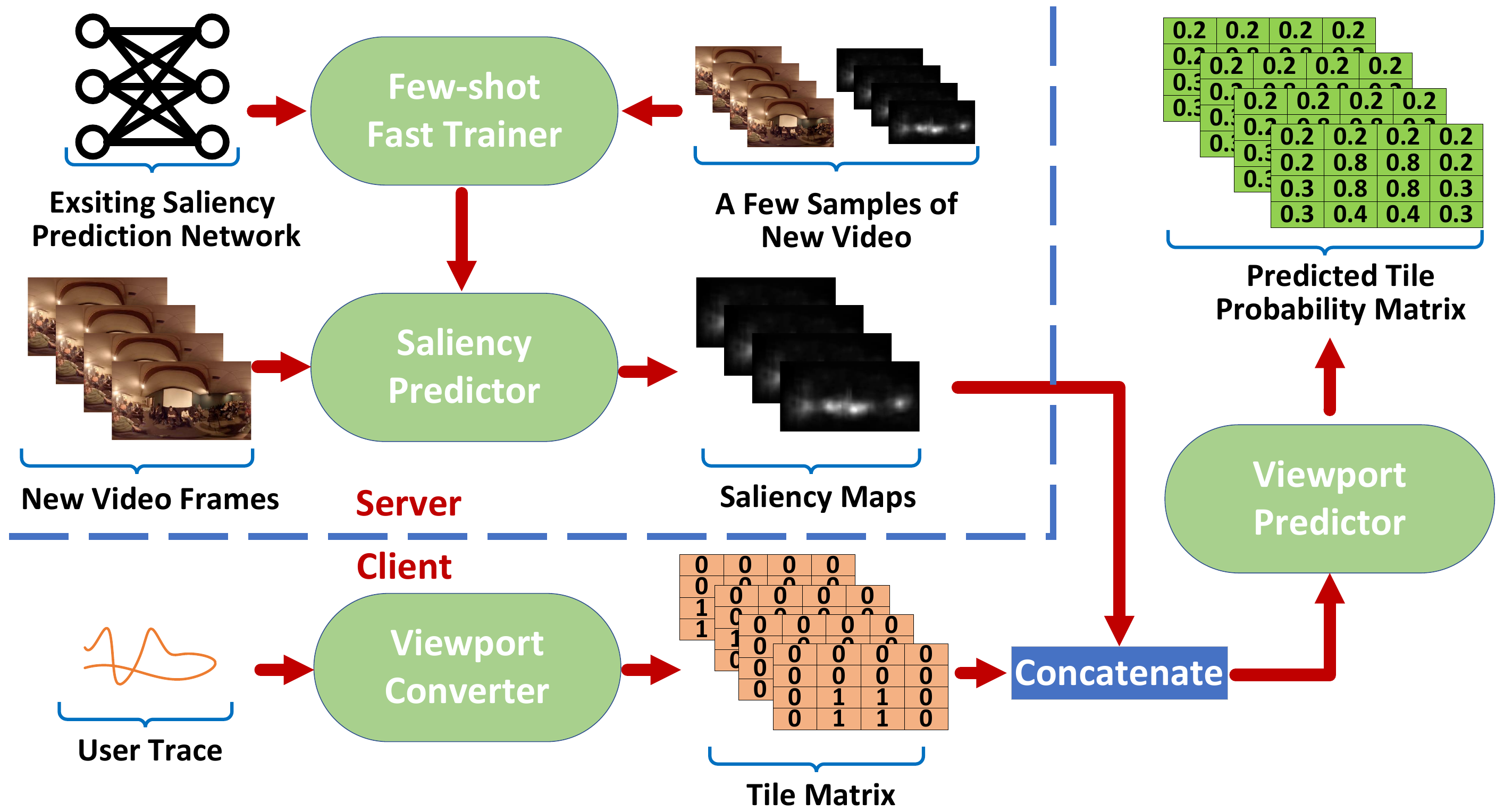}
	\caption{Overview of MFVP}
	\label{fig:archi}
\end{figure*}

To deal with the challenges, we propose an advanced learning-based Mobile-Friendly Viewport Prediction (MFVP) approach, which leverages both spatial and temporal information to make comprehensive viewport predictions and smartly distributes the workloads to introduce little computation overhead and transmission overhead to mobile devices. 
Fig.~\ref{fig:archi} shows the overview of our MFVP design. We first predict the saliency maps by using an advanced graph convolutional network (GCN) model, which helps learn how the interests of multiple users distribute over the objects spatially in the 360-degree video frames. 
In order to better meet the real-time requirements and complexity of live streaming’s content, we abandon the traditional pretrained method. We combine meta-learning and few-shot learning to rapidly train a saliency prediction network that can adapt to new tasks in time. In particular, we propose a MAML-based fast trainer of the prediction network. When a new video comes, our MAML-based trainer achieves the fast training from the previous prediction model using few new video chunks samples. Then the prediction model can be fine-tuned with new data. With our fast trainer, we successfully transfer the knowledge from the previous prediction model to the new videos’ model.
The saliency prediction results, together with the history of the user's viewport movement, are then fed into a modified long short-term memory (LSTM) model to extract the temporal patters from the series of data traces, which outputs the viewing probabilities of the tiles as the final future viewport prediction. The learning models are placed on the server side and the client side respectively, so that the computation workload on mobile client is acceptable and can be completed in real-time. As the prediction task is accomplished by the cooperation from both sides of the streaming, the communication overhead is managed to have negligible impact on bandwidth consumption. By formulating and solving a QoE optimization problem, we further investigate how to effectively integrate MFVP into a typical viewport-aware and rate-adaptive streaming system and design an efficient bitrate adaptation algorithm, which is evaluated together with MFVP for the 360-degree video live streaming performance.

In summary, this paper has the following contributions.
\begin{itemize}
	\item We propose MFVP, a learning-based viewport prediction scheme composed of an advanced GCN model and a modified LSTM model. The former is used for saliency prediction to output the image areas of user interests, while the latter is used to predict future viewport by leveraging the saliency maps and the historical viewports. 
	\item To make predictions in real-time and solve complex prediction tasks, we use MAML to rapidly train a saliency prediction network on each new video. Instead of using pretrained models, our MAML-based trainer solves complex tasks with little latency or overhead, resulting in excellent real-time performance. Based on our fast trainer, the saliency prediction network for every new video can be trained quickly with only 5 samples and 10 epochs.
	\item We implement MFVP with practical techniques and further enhance it by reducing the computation cost and the transmission cost to ensure that MFVP can run fast enough for live video streaming. 
	We carefully split the computation on both sides of the streaming and minimize its communication overhead by reducing data sampling frequency/transmission rate, downsizing the saliency maps, and compressing the learning models. 
	\item We showcase MFVP's application in a 360-degree video streaming system. We discuss how to integrate our viewport prediction approach with a typical bitrate adaptation scheme for tile-based viewport-adaptive streaming. The performance of the MFVP-supported 360-degree video adaptive streaming is evaluated to be superior.
\end{itemize}

The rest of the paper is organized in the following order. Section~\ref{sec:related} surveys related works on viewport prediction. 
We describe the design and implementation of the viewport prediction approach MFVP in Sections~\ref{sec:design} and~\ref{sec:implementation}, respectively.
The adaptive live streaming 
system
are presented in Section~\ref{sec:streaming}. In the end, we discuss results of experimental performance in Section~\ref{sec:experiment} and conclude this paper in Section~\ref{sec:conclusion}.

\begin{table}[t]
	
	\renewcommand{\arraystretch}{1.3}
	\caption{Comparison between our work and representative existing studies}
	\label{tab:related}
	\centering
	\begin{center}
	\scalebox{0.95}
	{
	\begin{tabular}{|c|c|c|c|}
		\hline
		\textbf{Methods}                                          & \textbf{Accuracy} & \textbf{\begin{tabular}[c]{@{}c@{}}Suitable for\\ Live Streaming\end{tabular}} & \textbf{\begin{tabular}[c]{@{}c@{}}Suitable\\ for Mobile\end{tabular}} \\ \hline
		LR \cite{qian2018flare}                   & 65$\%$$\sim$73$\%$               & \checkmark                                   & \checkmark                                     \\
		GCN \cite{lv2020salgcn}                   & 69$\%$$\sim$79$\%$               & ×                                                           & ×                                                             \\
		MobileNetV2 \cite{sandler2018mobilenetv2} & 74$\%$$\sim$83$\%$            & \checkmark                                   & \checkmark                                     \\ 
		LiveDeep \cite{feng2020livedeep}          & 77$\%$$\sim$86$\%$            & \checkmark                                   & ×                                                             \\
		MFVP (Ours)                                                      & 91$\%$$\sim$96$\%$              & \checkmark                                   & \checkmark          \\ \hline                          
	\end{tabular}
}
\end{center}
\vspace{0.1cm}
\end{table}

	\section{Related Work}
	\label{sec:related}

	Viewport prediction methods have been widely studied and employed in 360-degree video streaming.
	Existing systems usually use simple methods for viewport prediction. Flare~\cite{qian2018flare} is a practical system for streaming 360-degree videos on commodity mobile devices. It proposes a lightweight method for viewport prediction, which uses linear regression to predict the user's viewport trajectory. Rubiks~\cite{he2018rubiks} is a 360-degree streaming framework based on Android smartphones. It uses linear regression to predict head movement. This type of simple ML methods usually rely on the user behavior information. Nasrabadi et al.~\cite{nasrabadi2020viewport} proposed a cluster-based viewport prediction method, which classifies the users according to the head movement trajectories and assigns new users to the existing clusters to predict viewports. Based on KNN, Ban et al.~\cite{ban2018cub360} propose to predict the future viewport based on the user's personalized information and multi-user behavior information. 
	Damme et al.~\cite{van2022machine} propose a general, content-independent viewport prediction method that classifies behavior patterns based on user clustering and trajectory correlation.
	However, these methods only consider the user's historical viewport trajectory, and do not utilize the information of the video content.
	
	Xu et al.~\cite{xu2019analyzing} found that deep learning methods have better performance than simple methods, especially in the more challenging situations of longer prediction windows or more dynamic viewport movements. Fan et al.~\cite{fan2017fixation} proposed the fixation prediction networks and input HMD orientations, saliency maps, and motion maps into the LSTM network. Xu et al.~\cite{xu2018gaze} proposed to predict the viewport based on the history scan path and the image contents. It feeds the saliency map and the corresponding images to the CNN to extract features, uses LSTM to encode the history scan path, and then combines the CNN features and the LSTM features to predict the user's future viewport displacement. Nguyen et al.~\cite{nguyen2018your} proposed a head movement prediction framework based on the panoramic saliency and the head orientation history. It uses DCNN to learn saliency maps and uses LSTM network to predict viewports. \textcolor{black}{ Li et al.~\cite{li2019very} and Sun et al.~\cite{sun2022live} predict the user’s long-term viewport using a ConvLSTM-based encoder-decoder structure.}
	Although MobileNetV2\cite{sandler2018mobilenetv2} is a learning-based method that can be used on the mobile devices, it sacrifices accuracy for feasibility~\cite{bochkovskiy2020yolov4}.


	Using the traditional planar CNN model to process 360 video leads to a problem that the image projection will cause distortion and affect the prediction performance of the model. 
	Therefore, Wu et al.~\cite{wu2020spherical} proposes a 360-degree feature extraction network based on spherical CNN, combined with a recurrent neural network to extract users' personal preferences for 360-degree video content from viewing history for viewport prediction.
	SalGCN~\cite{lv2020salgcn} proposes a graph convolutional network (GCN) to predict saliency maps and use the high score area as the predicted viewport. 
	Zhang et al.~\cite{zhang2019drl360} designs a deep reinforcement learning-based 360-degree video streaming framework that uses LSTM to predict future viewports, but it doesn't take into account the video contents.
	PARIMA~\cite{chopra2021parima} leverages the fact that users are more likely to pay attention to the major objects in the video, so it first uses the object tracking algorithm to obtain the movement trajectory of the object and then collaborates with the user's historical viewport to predict the future viewport. Although these works can provide relatively accurate predictions, they cannot predict the viewport in real time. In order to predict the viewport in real time, LiveDeep~\cite{feng2020livedeep} employs the CNN model to analyze the video content and employs the LSTM model for the user's viewport trajectory. Its processing time for each video chunk was claimed to be smaller than the chunk duration, which thus satisfies the real-time requirement. Unfortunately, this approach cannot be supported on the mobile terminals. 
	LiveROI~\cite{feng2021liveroi} can be used for real-time VR streaming media, which uses an action recognition algorithm to analyze the video content, and uses the analysis results as the basis for viewport prediction. To eliminate the need for historical video/user data, LiveROI employs adaptive user preference modeling and word embeddings to dynamically select video viewports at runtime based on user head orientation. LiveObj~\cite{feng2021liveobj} develops a real-time viewport prediction mechanism to detect objects in videos based on semantics, and then track the detected objects using reinforcement learning algorithms to infer the user’s viewport in real-time.

	To summarize the existing viewport prediction methods, complex methods can obtain higher prediction accuracy, but are difficult to deploy in a lightweight manner for mobile devices, while simple methods may be able to apply on the mobile terminals, but the prediction performance is worse. Therefore, in this work, we attempt to maintain the merit of complex method to fully utilize available information and design a lightweight framework for mobile live streaming. Specifically, we propose a viewport prediction method based on saliency prediction, and separate the saliency prediction module from the user behavior prediction module to achieve a mobile-friendly deployment. We compare our work with the representative existing prediction approaches in Tab.~\ref{tab:related} to show the advantages of our MFVP approach.

\section{MFVP Design}
\label{sec:design}
In this section, we present the detailed design of MFVP.
\subsection{Overview}
Our MFVP has three modules that are responsible for fast network training, saliency prediction and viewport prediction, respectively. Fig.~\ref{fig:archi} illustrates MFVP's workflow. 
To better cope with complicated video content and to meet real-time requirements, we use the few-shot fast trainer to train a general network that easily adapts to the saliency prediction network for each video. We reuse finished live streaming data and quickly transfer knowledge from the previous model to new models.
After a quick training, 
the saliency prediction module takes raw 360-degree video frames as input and generates the corresponding saliency maps, which are the heat-maps for the interests of massive users. 
Taking advantage of the previous step, the viewport prediction module further considers the individual user's historical viewports together with the likelihoods of viewing different areas and outputs the final prediction. 
The network fast trainer and the saliency prediction runs on the server side, while the viewport prediction runs on the client side, which significantly reduces the computation on the mobile devices. 
Our design allows MFVP to make highly accurate viewport predictions by fully exploiting the strength of neural networks and learning from both the video frames (imagery data) and the history of user behavior (time series).
More importantly, as the computation and the communication are carefully managed across the server and the client, MFVP is feasible to run fast enough for live streaming on the less powerful mobile devices.

\subsection{Saliency Prediction}

In order to correlate image areas with different degrees of user visual preferences, convolutional neural networks (CNNs) are usually used and optimized for 2D frames in traditional videos~\cite{kummerer2014deep}~\cite{pan2016shallow}~\cite{zhang2016seed}. 
However, 360-degree videos provide spherical views. Directly applying traditional convolution kernel with a grid design to handle spherical features is clearly inappropriate.
Moreover, transforming spherical views into conventional 2D frames requires projection. 
As the most adopted projection method, equirectangular projection (ERP)~\cite{salomon2006transformations} is known to have distortions. 
Therefore, it is sub-optimal to use regular CNNs in saliency prediction for 360-degree videos. 
To deal with the unique feature of spherical views in 360-degree videos, we utilize the advanced graph convolutional network (GCN) model to extract visual features directly from spherical views. 

\subsection{Few-shot Fast Trainer for Saliency Prediction Networks}
In order to better meet the real-time requirements and the complexity of live video content, we propose a few-shot fast trainer for the saliency prediction network based on meta-learning, 
where the goal of meta-learning is to train a model on various learning tasks so that it can solve new learning tasks with a small amount of data samples.
Since the live video content cannot be known in advance, the prediction network should be able to handle the prediction tasks of different videos.
Considering the complexity of video content, the model needs to be fine tuned according to different videos for more accurate inference. 
\textcolor{black}{Due to the diversity of video content, we need to adjust the parameters of the model to adapt to different video content. The pre-training method based on video content may experience a sharp drop in performance after the video content is switched, making it unsuitable for live broadcast scenarios. The meta-learning module can help our model quickly converge and adapt to new video content. Fig~\ref{fig:MAML_result} shows the loss change curves in two different ways. It can be seen that our model can be fine-tuned faster using the meta-learning method. Therefore,} 
we use the meta-learning based few-shot fast trainer to quickly train the prediction model of each new video.
In this way, we can learn a model that can quickly adapt to new tasks with only a few samples, thus meeting the real-time requirements of live 360-degree video streaming.

\subsection{Viewport Prediction}
We design our viewport predictor to work in tile-based 360-degree video streaming. Instead of predicting the user's gaze point, our goal is to predict the likelihood that each tile will appear in the viewport. As viewport predictor should leverage as much information about video watching as possible, we process the viewport information so that it can be handled together with the saliency map. We map the trace of the user head movement into the tile grid, and use a 0-1 matrix to denote a viewport in a tiled frame, where 1 indicates the corresponding tile is watched and 0 indicates otherwise. In this way, we encode the location of the viewport in an image, which can be resized into the same format as the saliency map.
Then the viewport predictor predicts future viewports based on the historical viewports and the corresponding saliency maps, which utilize both temporal and spatial information.
\section{MFVP Implementation}
\label{sec:implementation}


In this section, we present the detailed implementation of MFVP and further discuss the enhancements for reducing the overhead for mobile clients.


\subsection{Saliency Prediction}
\label{sec:sp}
We adopt the state-of-the-art SalGCN method~\cite{lv2020salgcn} to predict saliency maps for 360-degree video frames. 
It mainly consists of three steps. First, convert ERP format frames into spherical image data and use the Geodesic ICOsahedral Pixelation (GICOPix)~\cite{yang2020rotation} to construct a spherical undirected and connected graph to sample the spherical image and collect the training data. Second, build a graph convolutional network for saliency prediction, which uses an encoder-decoder structure similar to U-net~\cite{ronneberger2015u} and uses the chebnet~\cite{defferrard2016convolutional} as the graph convolution layer to approximate the convolution kernel by recursively calculating the chebyshev polynomial. Third, output the spherical saliency prediction results from the trained GCN model in the format of the constructed graph, which are further transformed into the final saliency maps in the ERP format.

\begin{figure}
	\centering
	\includegraphics[width=0.99\linewidth]{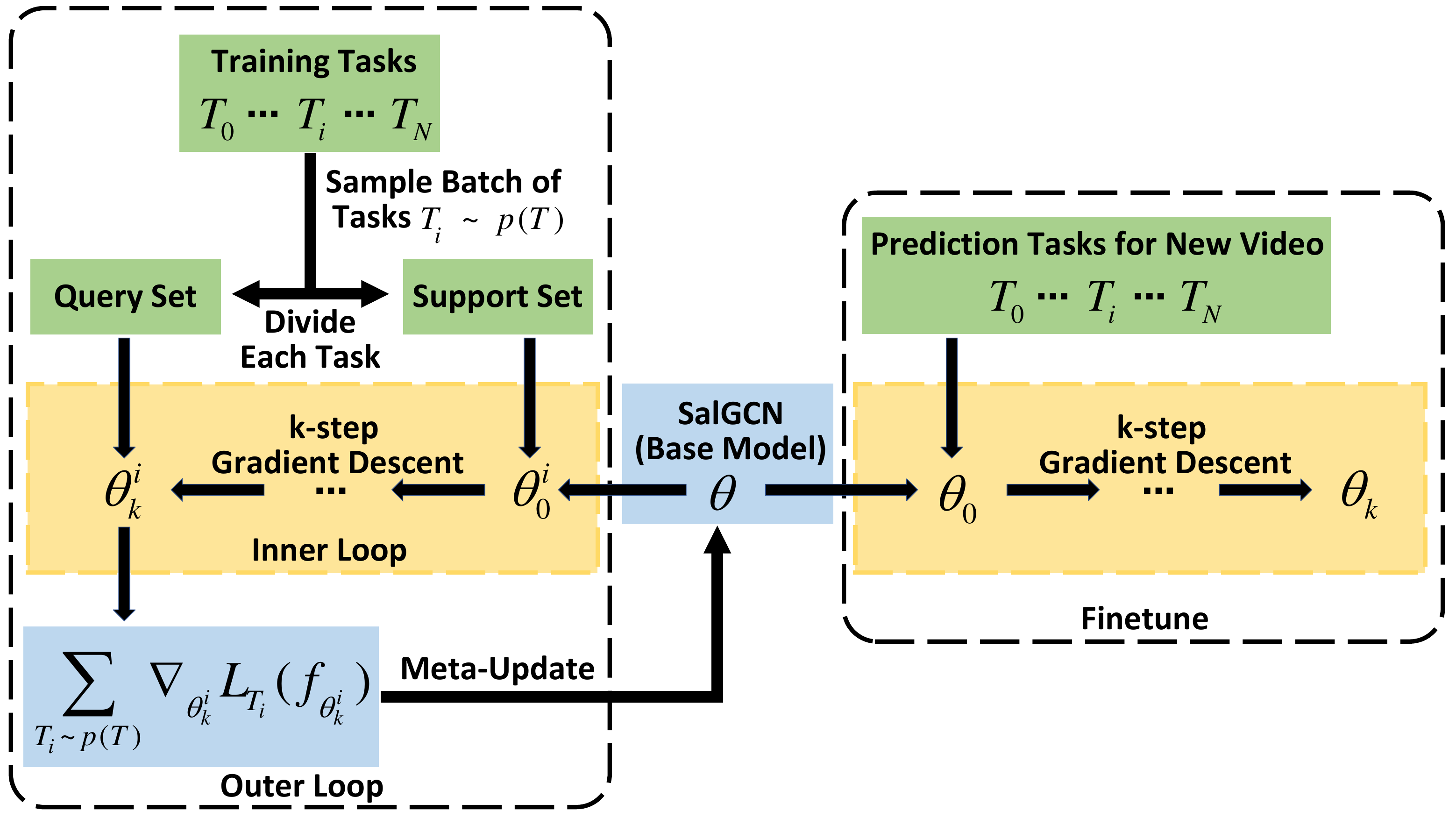}
	\caption{The Workflow of MAML-based Fast Trainer}
	\label{MAML}
	\vspace{-0.3 cm}
\end{figure}

\subsection{Few-shot Fast Trainer for Saliency Prediction Networks}
\label{sec:maml}
\textcolor{black}{Since the adaption effect of the pre-trained model to new videos is not ideal, we need a fast convergence method to adapt to new videos to meet the needs of live broadcast scenarios. At the same time, in the experimental part, Fig.~\ref{fig:MAML_result} also shows the loss curve of our method compared with the pre-training method.}  We use the model-agnostic meta-learning algorithm (MAML)~\cite{finn2017model} based trainer to quickly train the saliency prediction model of each new video.
We use MAML to learn good weight initialization for our prediction network that can adapt quickly to new videos with only a few gradient updates.
The parameter initialization is learned from many tasks, and it can be broadly applicable to various tasks. Therefore, we can quickly transfer the previous prediction network to the new task. In this way, we can learn a model that adapts to a new task with only a few samples and trains quickly in a few iterations without overfitting.
The few samples for fast training can be generated using a subset of users' viewing data.
\textcolor{black}{The process of generating the saliency map is performed on the server. After receiving the video clips, the inference of the saliency map can be performed, and we Model inference requires very little time. Because the video content changes, all videos are usually required to generate saliency. However, we use the meta-learning based method MAML, which can quickly adapt to new video content, so there is no need to obtain all videos. Fragments for saliency map generation.  The generated saliency map is relatively small data of 10242 points, and the required transmission overhead is very small and can be ignored.  Therefore, our method can well solve the prediction problem and bit allocation problem of real-time systems.}

Formally, a dataset defines a distribution $p(T)$ for a series of saliency prediction tasks.
Each task $T_i$ is sampled from the distribution $p(T)$ and divided into a support set and a query set.
The goal of the fast trainer is to learn a general model $f_\theta$ with parameters $\theta$ that can be quickly adapted to new tasks in the distribution.

As shown in figure~\ref{MAML} , our fast trainer consists of two optimization loops and a fast fine-tuning stage. \textcolor{black}{For MAML training of new videos, after the new video comes, the model will be continuously updated with the content of the new video and the MAML method. At the same time, the predictions at this time are carried out simultaneously, and predictions can still be made continuously. With the rapid convergence of MAML, we can quickly adapt to new videos and video scene switching.}
The outer loop updates parameter initialization so that the model can quickly adapt to new tasks. 
At each iteration of the outer loop, we sample a batch of saliency prediction tasks.
For each task $T_i$ from the batch, 
the inner loop performs adaptation on the support set using the base model $f_\theta$ from the outer loop.
After $k$ steps gradient descent on $T_i$, the parameters become $\theta_k^i$, which can be expressed as:
\begin{align}	
	\theta_k^i=\theta_{k-1}^i-\alpha\nabla_{\theta_{k-1}^i}L_{T_i}(f_{\theta_{k-1}^i}), 
	\label{eq:theta_k}
\end{align}
where $\alpha$ is the learning rate, $L_{T_i}$ is the loss function of the task $T_i$, which is the KL loss with sparse consistency proposed by SalGCN~\cite{lv2020salgcn}.
Then calculate the loss of the adapted model on the corresponding query set.
The outer loop performs meta-updates on $\theta$, minimizing the total loss on the query set of tasks in the batch.
The meta-update of the basic model parameters $\theta$ can be expressed as follows:
\begin{align}	
	\theta\gets\theta-\beta\nabla_{\theta}\sum_{T_i\sim p(T)}L_{T_i}(f_{\theta_{k}^i}), 
	\label{eq:theta}
\end{align}
where $\beta$ is the meta learning rate. To reduce the overhead of computational resources, we use the first-order approximation of MAML, ignoring the second-order derivatives. The first-order approximation can reduce a lot of time and memory usage without significantly degrading performance, using the following updates:
\begin{align}	
	\theta\gets\theta-\beta\sum_{T_i\sim p(T)}\nabla_{\theta_{k}^i}L_{T_i}(f_{\theta_{k}^i}). 
	\label{eq:theta_fomaml}
\end{align}
Finally, in the fast fine-tuning stage, the predictive model is initialized with the parameters trained with MAML. After that, the model only needs to use a small number of target video samples to obtain a relatively good prediction ability for new videos after several rounds of gradient updates.

Our fast trainer enables knowledge transfer from existing predictive models to new predictive models. 
Our fast trainer greatly reduces the need for training data and speeds up the training process. 
It ensures that our system responds quickly and only a few learning iterations with a small number of samples are needed to obtain an acceptable model.
Furthermore, our model can be fine-tuned whenever new video data and user viewing data become available.
Therefore, our fast MAML-based trainer is essential when applied to live 360-degree video streaming, as our prediction network has the ability to adapt quickly to meet the real-time requirements of performing saliency prediction and subsequent viewport prediction in live streaming. 

\subsection{Viewport Prediction}
\label{sec:vp}
In order to learn patters from the sequential images, we adopt the convolutional LSTM (ConvLSTM)~\cite{xingjian2015convolutional}, which utilize both spatial and temporal information.
And we modify it according to the needs of live streaming.
Standard ConvLSTM, as a convolutional counterpart of conventional fully connected LSTM (FC-LSTM)~\cite{hochreiter1997long}, introduces convolution operation into input-to-state and state-to-state transitions.
ConvLSTM preserves spatial information as well as modeling temporal dependency. Thus it has been well applied in many spatiotemporal tasks, such as dynamic visual attention prediction~\cite{wang2018revisiting}, video super-resolution~\cite{guo2017building}.

However, the standard ConvLSTM has a relatively large computational cost and memory consumption. We need to reduce the computational cost and parameter amount of ConvLSTM to meet the requirements of mobile computing and live 360-degree video streaming.
The computation of convolution layers in most models accounts for a majority of FLOPs and running time~\cite{lu2019augur}. 
Inspired by the popular acceleration techniques of standard convolution layers~\cite{szegedy2016rethinking}\cite{howard2017mobilenets}\cite{li2019very}\cite{sun2022live}, we replace each convolution in the ConvLSTM cell with a depthwise separable convolution~\cite{chollet2017xception}, 
which greatly reduces the computation time and the number of parameters of ConvLSTM with little loss of accuracy.
In this case, each input channel is independently convolved with one filter (called depthwise convolution) and a 1×1 (pointwise) convolution is applied after the depthwise convolution to combine the outputs of the depthwise convolution layers.
In addition, the saliency map and tile matrix have different effects on the prediction results, that is, the importance of different channel features is different. In order to learn the importance of different channel features for better prediction, we introduce the Squeeze-and-Excitation (SE) block~\cite{hu2018squeeze} into ConvLSTM.
Although the convolution operation implicitly introduces weights for every channel, these implicit weights are not specialized for every image. To explicitly import weight on each network layer for each image, we extend each depthwise separable convolution layer with SE block, which computes normalized weight for every channel of each item. By multiplying weight learned by SE block, feature maps computed by convolution are re-weighted explicitly.
Because SE block is lightweight, it introduces only a small amount of computation time and number of parameters.

\begin{figure}
	\centering
	\includegraphics[width=0.95\linewidth]{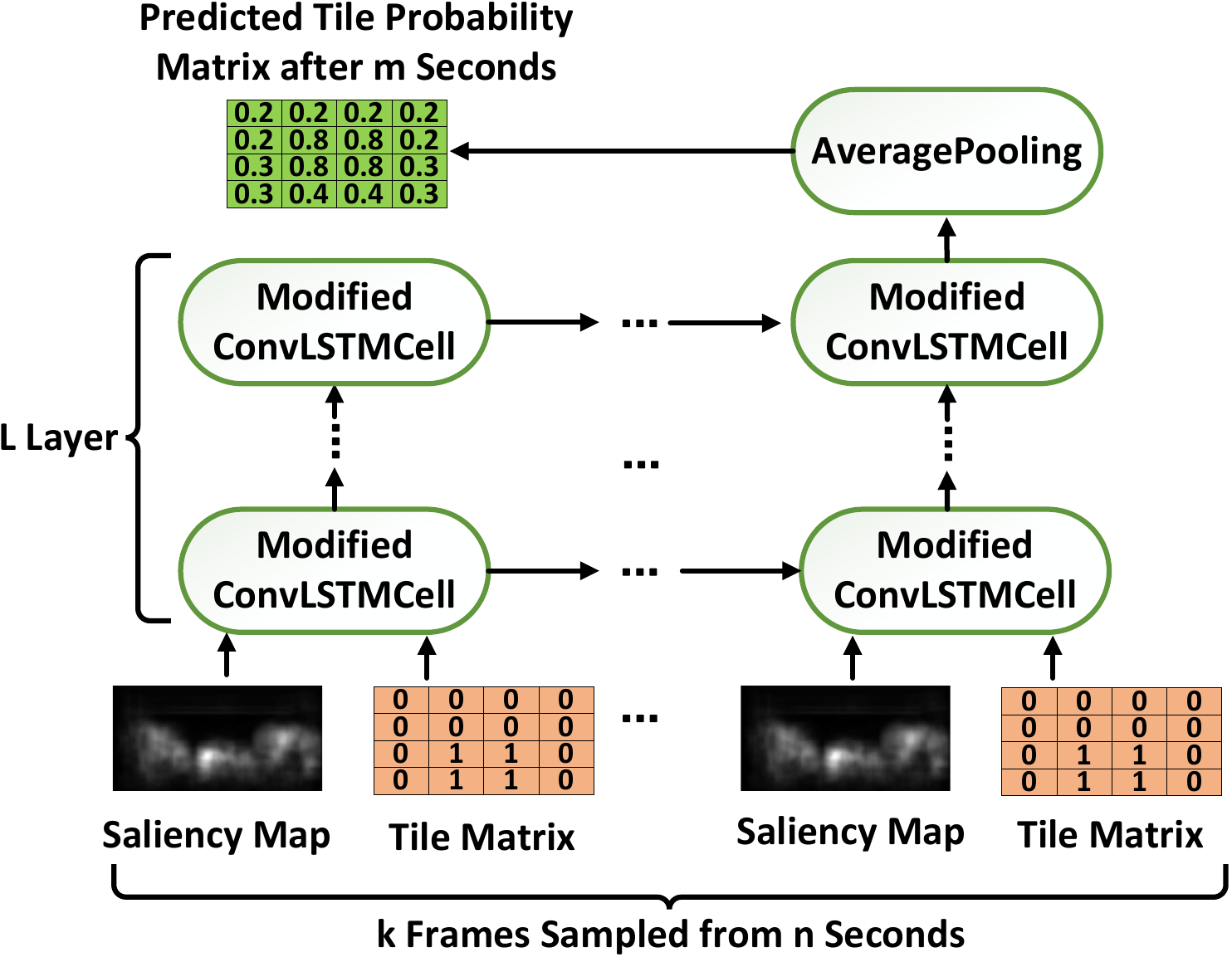}
	\caption{Structure of ConvLSTM model}
	\label{fig:lstm}
	\vspace{-0.3 cm}
\end{figure}

\begin{figure}
	\centering
	\includegraphics[width=0.95\linewidth]{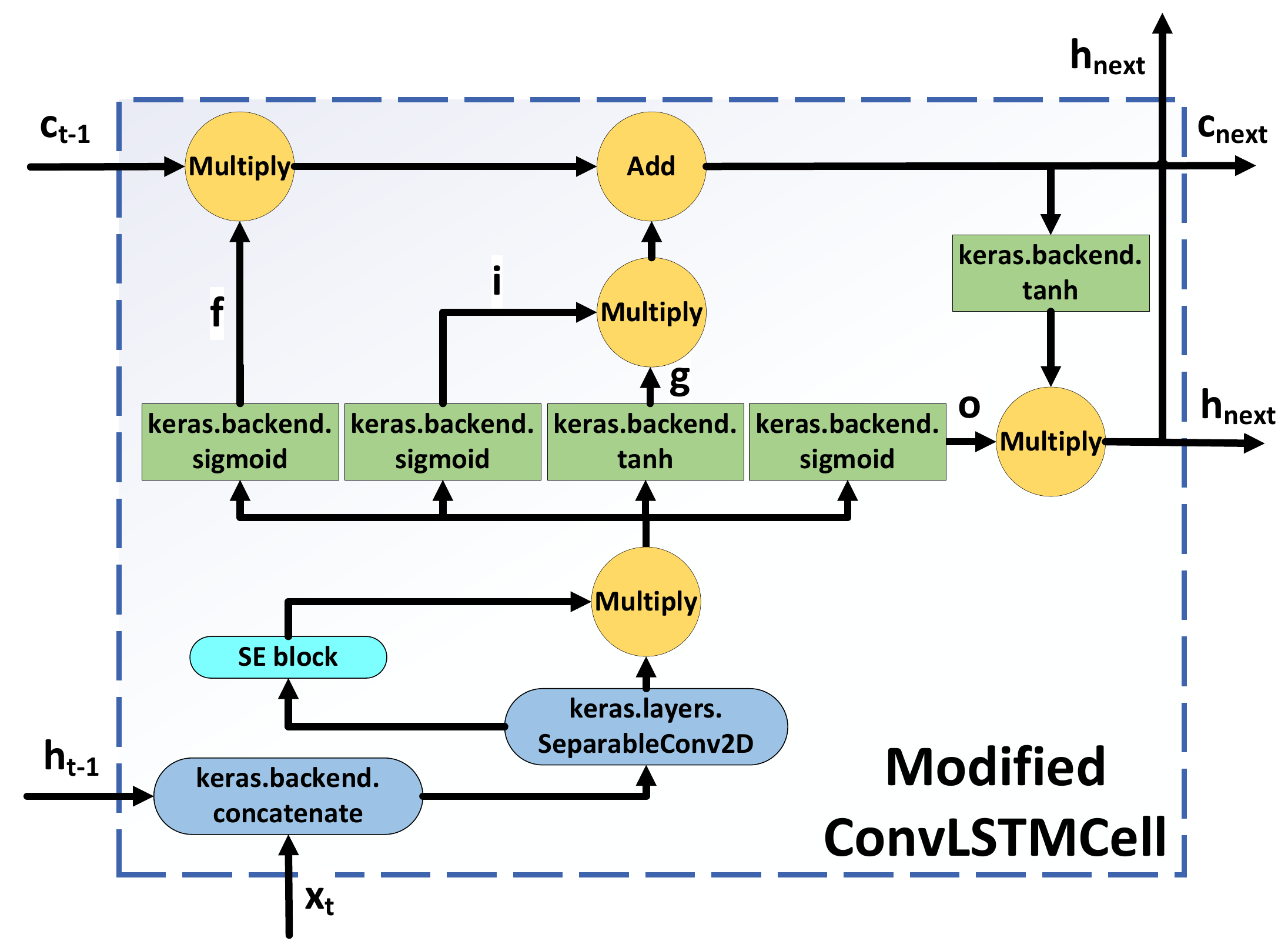}
	\caption{Structure of Modified ConvLSTMCell}
	\label{fig:cell}
	\vspace{-0.3 cm}
\end{figure}

A general solution to obtain a deep learning model that works on mobile devices is to train the model with the server-side frameworks (e.g., TensorFlow, Pytorch~\cite{pytorch}, and Caffe2~\cite{caffe2}) and then transform it into a mobile supported version using the development tool (typically, TensorFlow Lite). Unfortunately, TensorFlow Lite defines a very limited number of operators and thus only supports the most basic calculations and operations. Naively using the \textit{convlstm} interface of Keras~\cite{keras} will fail due to the unsupported operators in the mobile learning framework. 
To this end, we implement modified ConvLSTM using the mobile-supported basic operators with Keras.

\begin{figure*}
	\centering
	\begin{minipage}{0.24\linewidth}
		\centering
		\includegraphics[width=1.0\linewidth]{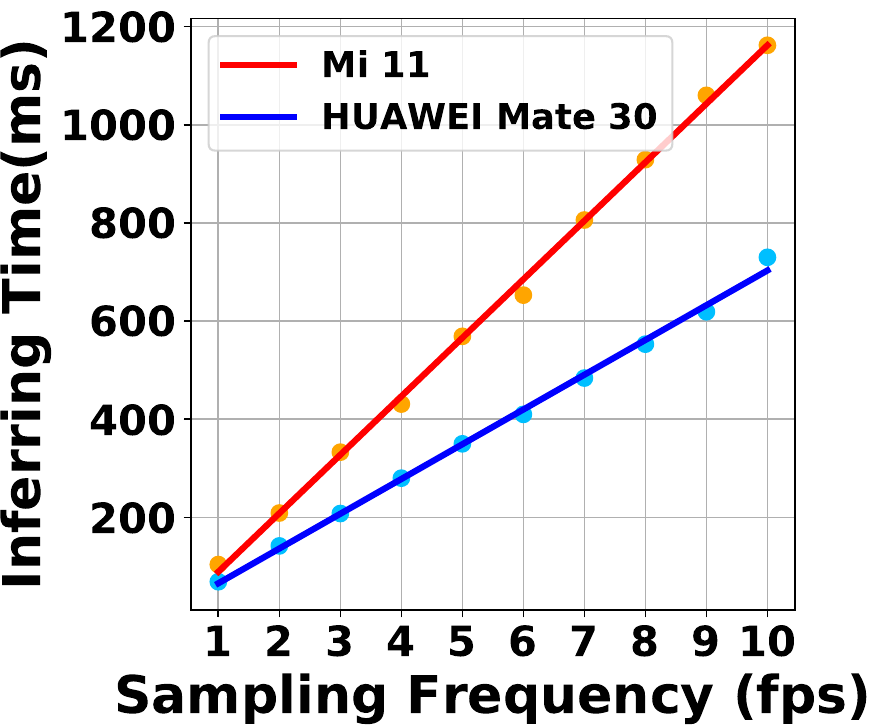}
		\caption{\small Inferring time \textit{vs.} Sampling frequencies}
		\label{sf-time}
	\end{minipage}
	\begin{minipage}{0.24\linewidth}
		\centering
		\includegraphics[width=1.0\linewidth]{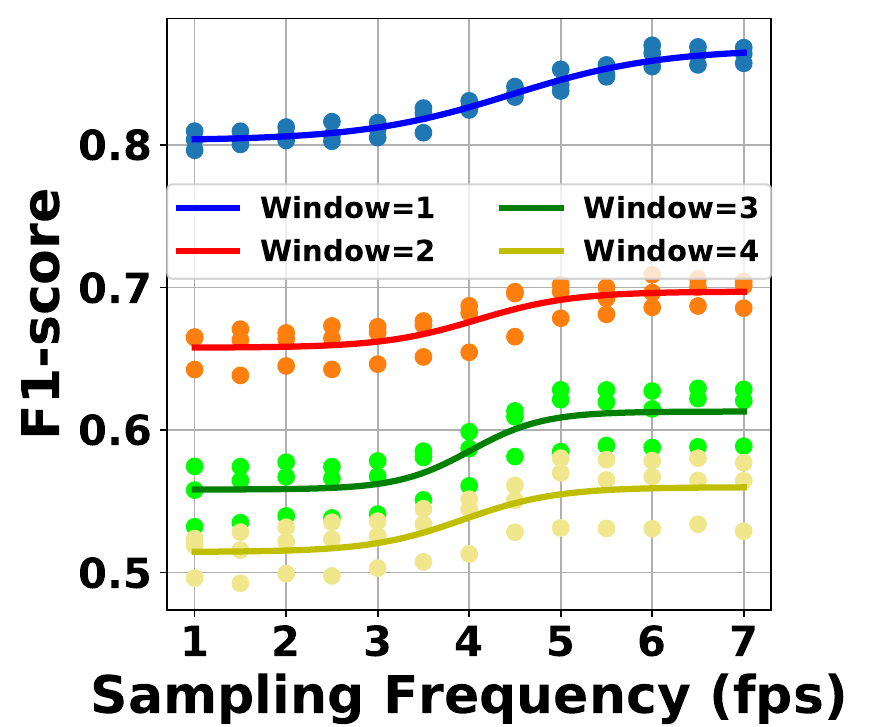}
		\caption{\small Prediction performance \textit{vs.} Sampling frequencies}
		\label{fig:sf-f1}
	\end{minipage}
	\begin{minipage}{0.24\linewidth}
		\centering
		\includegraphics[width=1.0\linewidth]{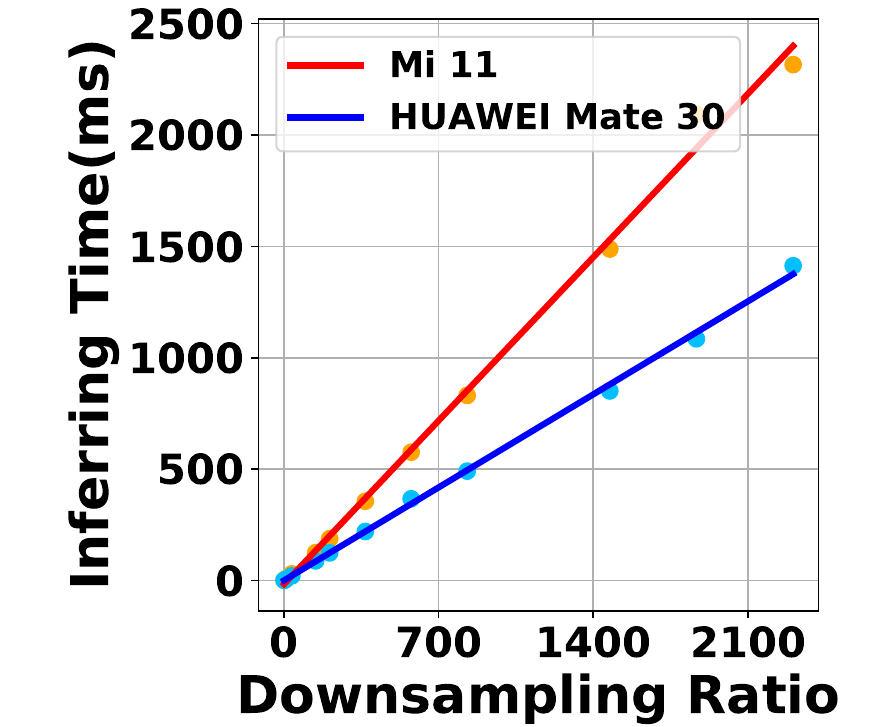}
		\caption{\small Inferring time \textit{vs.} Downsampling ratios}
		\label{dr-time}
	\end{minipage}
	\begin{minipage}{0.24\linewidth}
		\centering
		\includegraphics[width=1.0\linewidth]{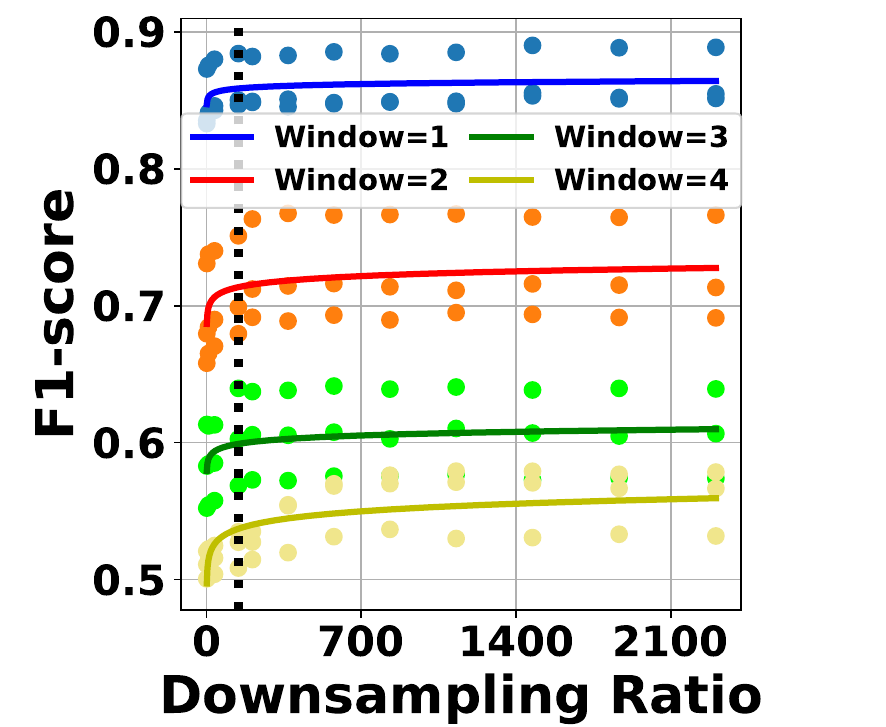}
		\caption{\small Prediction performance \textit{vs.} Downsampling ratios}
		\label{fig:ratio-f1}
	\end{minipage}
\end{figure*}

The first step is to build modified ConvLSTMCell, as shown in Fig.~\ref{fig:cell}. 
We use keras.backend.concatenate to concatenate the current input $x_t$ and the previous hidden state $h_{t-1}$, 
and then use keras.layers.SeparableConv2D to convolve it and multiply by the weight of the SE block output to get three gated states (i.e. a forget gate $f$, a select memory gate $i$, and an output gate $o$) and one input data $g$, respectively. 
Forget gate $f$ indicates which of the last long-term memory cell $c_{t-1}$ needs to be retained and which needs to be forgotten. Select memory gate $i$ selectively memorizes the input $x_t$. Output gate $o$ indicates which ones will be output as the current state. Specefically, $i$, $f$, and $o$ are converted to values between 0 and 1 through the keras.backend.sigmoid activation function, and $g$ is converted to values between -1 and 1 through the keras.backend.tanh activation function. Finally, use the four states and the previous long-term memory $c_{t-1}$ to calculate the next long-term state $c_{next}$ and the next hidden state $h_{next}$, which can be calculated as follows: 
$c_{next}=f*c_{t-1}+i*g$ and $h_{next}=o*\textit{keras.backend.tanh}(c_{next})$, 
where '$*$' denotes the Hadamard product.
%
All the gates $i$, $f$, $o$, $g$, the long-term memory cell $c_{t-1}$, $c_{next}$ and the hidden state $h_{t-1}$, $h_{next}$ are 3D tensors.
We then connect multiple modified ConvLSTMCells into a ConvLSTM and finally add an Average Pooling layer and sigmoid activation function. The loss function uses binary cross entropy, and the optimizer uses RMSprop. 
Fig.~\ref{fig:lstm} shows the structure of our ConvLSTM model.
We concatenate the saliency map and the tile matrix for historical user viewport as the input of ConvLSTM. \textcolor{black}{ Based on previous work~\cite{lo2017360}, this tile is usually a 10X20 matrix. The height is evenly divided into 10 parts, and the width is evenly divided into 20 parts, resulting in 200 tiles. For the tile matrix, the Tile labeling in ViewPort is one, and the remaining Tile labels are zero.} The probability matrix for each tile to appear in the viewport is output as the prediction result, which can be further used for bitrate adaptation in the 360-degree video streaming.
After training our redesigned ConvLSTM model on the server side, we use TFLiteConverter to convert it into a TensorFlow Lite model that can be deployed on the mobile terminal.

\subsection{Reducing Overhead for Mobile Clients}
\label{sec:overhead}

We next analyze the overhead introduced by MFVP approach and investigate to improve its efficiency without undermining the feasibility.




\subsubsection{Computation Workload on Mobile}
Our first design for computation overhead is placing the deep learning model on server and on client respectively, so that the computation tasks of MFVP can be carefully completed by both sides of the streaming. As the 360-degree videos and the historical watch events from different users can be accessed from the service provider, MFVP's model training can be mostly done on the server offline. MFVP places the saliency prediction on the server because it requires no engagement from the user during the video watch event and can be done ahead of the user's video streaming. Moreover, MFVP runs the viewport prediction on the client, where the historical viewports can be recorded for viewport prediction by the mobile devices and the prediction results are further referenced for making rate decisions.



Using neural networks (NNs) to make predictions inevitably introduces considerable computation on mobile clients, which have limited computation power for NN-based inference and insufficient memory to load the model. A widely used technique is to compress the NN model and examine the accuracy-complexity trade-off. Fortunately, the hardness of our viewport prediction task varies. When the 360-degree video frame is split into less tiles, the viewport is fitted at a coarser grain and thus can be more easily predicted. We can adjust the number of layers in the ConvLSTM model of MFVP according to the level of viewport prediction task's difficulty. \textcolor{black}{As for whether this means that the model structure will be adjusted for each new real-time session according to the difficulty level of prediction, this is currently only theoretically adjustable according to the difficulty of prediction. For example, if the tile segmentation granularity is coarser, the prediction task will be simpler. It is naturally better to predict, and the number of convlstm layers does not need to be so many.
Similarly, even within a single session, if we find that the user's head movement trend is consistent or slow, we can use a rougher prediction. However, there is currently no such mechanism that can automatically adjust. This article only explains a possibility in this regard. In future work, we will consider in-depth thinking and experimental verification in this direction.}

It should be noted that the prediction results are not reported frame by frame. In a typical video streaming system, the video data are requested chunk by chunk, and thus the viewport prediction should also be reported chunk by chunk to assist the bitrate adaptation. Given the viewport prediction task of a specific difficulty, the total running time of viewport prediction for each chunk depends on how many tasks are executed. We can vary the sampling frequency for the user historical head movement to control the arrival rate of prediction tasks. On one hand, higher sampling frequency for user behavior will introduce more prediction tasks and thus increase the required execution time. In a live streaming scenario, we need to keep the viewport inferring time less than the chunk length, so that the prediction results can be obtained in time. 
On the other hand, if we reduce the sampling frequency too much, there is not enough training data to make accurate viewport predictions.
We conduct real-world measurements on two test mobile devices (Xiaomi Mi 11 and Huawei Mate 30) and plot the viewport inferring time for each chunk under different sampling frequencies in Fig.~\ref{sf-time}. A direct observation is that the inferring time increases linearly with the rising sampling frequency for user behavior.
As shown in Fig.~\ref{fig:sf-f1}, we also check the corresponding prediction performance in terms of F1 score. The results suggest that we need to set the sampling frequency to an appropriate value so that the computation overhead is reduced while the prediction performance is kept at a good level.

As the size of saliency maps and the tile matrix needs to be coordinated into the same size input for ConvLSTM, the effective resolution after resizing the saliency maps directly affects the amount of computation. We define the downsampling ratio as the ratio between the pixel count of the downsized saliency map and the number of tiles in a frame, which reflects the degree of the saliency map downsampling. We measure the viewport inferring time with different downsampling ratios in Fig.~\ref{dr-time}, which roughly exhibits linear increasing patterns for the both test devices. By exploiting this linear function, we can estimate the computation workload when the downsampling ratio is properly adjusted.


%
%

\subsubsection{Communication Overhead}
MFVP introduces two types of communication overhead: the one-time transmission for the ConvLSTM model and the continuous transmission for the saliency maps during the streaming. 
First, to reduce the overhead of model transmission, we conduct a post-training float16 quantization~\cite{float16} for this model running on mobile devices, which converts the model weights to 16-bit floating point values from the original float32. It achieves 2x reduction in model size (down to 710 KB in our case) with negligible accuracy loss. We further use file compression technique to compress the model in zip format, which further reduces the model size to only 10 KB. 

Second, to reduce overhead of saliency map transmission, MFVP approach only sends necessary saliency maps, which can be selected for further compression. 
We find that the saliency mayps for the same video chunk are very similar, and thus they can be send at a very low rate (one
saliency map per chunk in our case) without noticeably hurting the viewport prediction performance. MFVP thus avoids unnecessary transmission for saliency maps by identifying the client's actual demand. We also observe that the saliency maps does not have to be in very high resolutions, which can be downsampled with much less pixels while maintaining similar prediction performance.
We check the viewport prediction performance under different downsampling ratios with various prediction window sizes in Fig.~\ref{fig:ratio-f1}, where the curves can be fitted in the form of $F1\_score=b*ratio^{-a}+c$. The result confirms that we can reasonably downsize the selected saliency maps to reduce communication overhead while maintaining good prediction accuracy.



\section{Adaptive Live Streaming for 360-degree Videos}
\label{sec:streaming}

In this section, we discuss the application of our MFVP in a typical 360-degree video adaptive streaming system.

\begin{figure}
	\centering
	\includegraphics[width=0.95\linewidth]{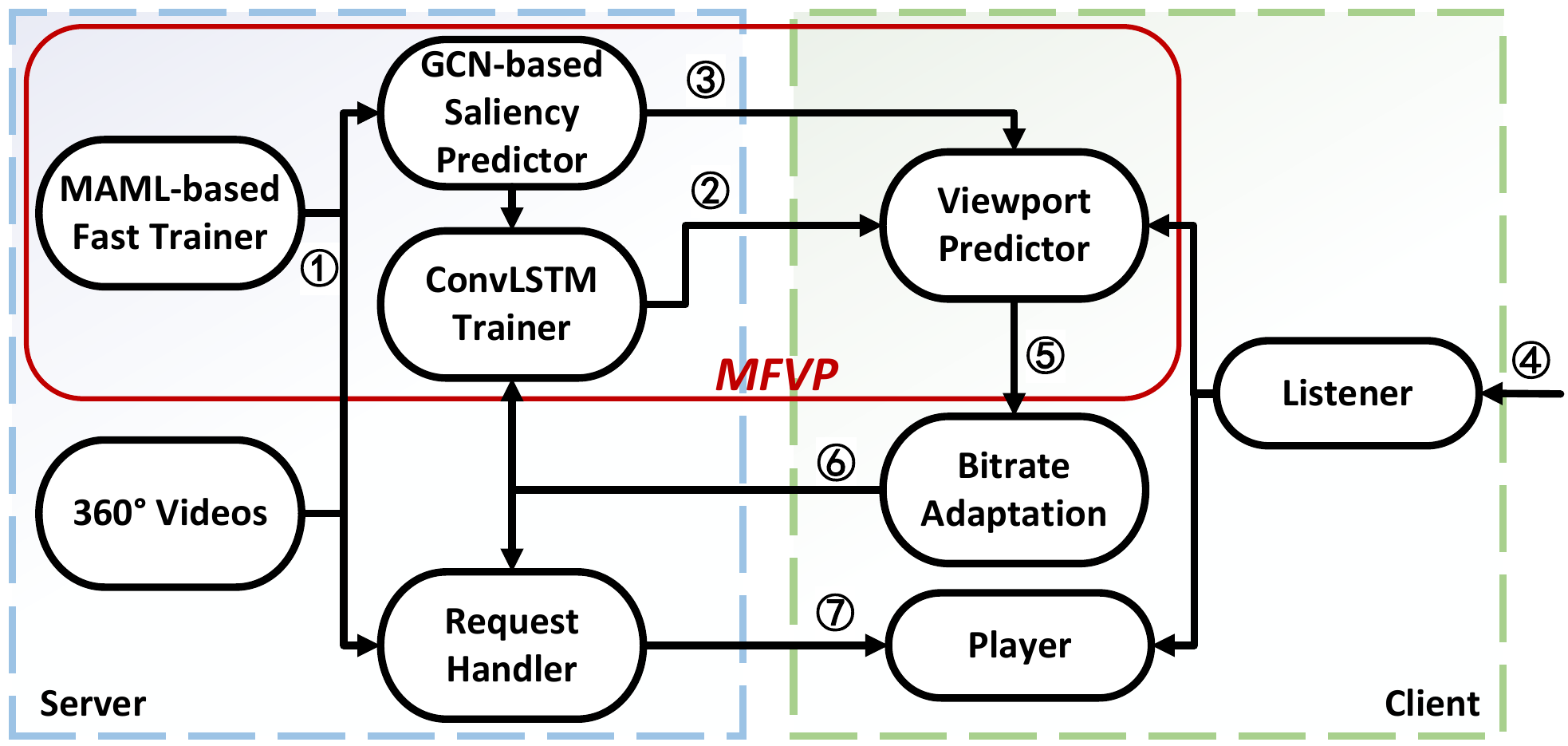}
	\caption{Architecture of the 360-degree video streaming system}
	\label{fig:system}
	\vspace{-0.3 cm}
\end{figure}

\subsection{System Architecture}
We present the system architecture in Fig.~\ref{fig:system}, which integrates MFVP and the adaptive streaming service. The streaming system works as follows. 
(1) The server trains the SalGCN model using MAML-based fast trainer and trains the ConvLSTM model using the stored 360-degree videos and the data collected from tile requests in past video watch events. 
When a new video is cached on the server, the saliency maps are generated using the pre-trained SalGCN model. 
(2) Whenever a user initiates a streaming session, the server check the cookies to see if s/he is a new user. If the user is new, it sends to the client the ConvLSTM model pre-trained using the traces from other users who watch the same video; if not, it sends the ConvLSTM model that is trained incrementally by each time the user watches a video. Note that the model is compressed for this one-time transmission. 
(3) Along with the requested tiles for each video chunk, the server sends back the saliency maps for the chunk at a certain rate. 
Upon receiving the saliency maps, the mobile client is able to make viewport predictions using the ConvLSTM model. 
(4) It listens for user interactions to obtain the viewport location at a specific frequency, which information is used to output the probability matrix for the tiles to be watched. 
(5) With the viewing probability matrix, the bitrate adaptation algorithm can make bitrate decisions.
(6) Appropriate requests are made by the client for tiles with different resolutions to the request handler on the server. 
(7) The requested tiles for the future chunks are transmitted, buffered, and then played in the client's player. The related information is collected during the video watch event for the incremental model training.

The bitrate adaptation algorithm is the core component of an adaptive live streaming system. Its goal is to assign an appropriate bitrate to each tile given the available bandwidth to maximize the user experience in terms of the video quality and the quality churn in the viewport. 
The bitrate adaptation is highly dependent on the prediction result of MFVP, while other modules in the streaming system are relatively independent, so we next propose a bitrate adaptation algorithm based on MFVP.

\subsection{Bitrate Adaptation Using MFVP}
We next present the formulation of bitrate adaptation optimization problem for 360-degree video live streaming and propose our solution design. 

\subsubsection{Problem Formulation}
We consider a live streaming system, where the 360 video is streamed chunk by chunk with a fixed chunk length $L$. 
Each chunk is divided into $m*n$ tiles and $p_{i,j}$ denotes the viewing probability of tile $(i,j)$ in row $i$ and column $j$ which is reported by MFVP approach. 
Let $SF$ be the sampling frequency for user behavior on the mobile client.
Once MFVP approach has made the viewport prediction, the result can be utilized right away in the bitrate decision-making for the predicted chunk. 

Because the available bitrate levels are limited, tiles do not need to be distinguished too finely by probability, but can be treated as different classes. 
Therefore, we propose a classification-based bitrate adaptation scheme (CBA), which classifies tiles according to the distance from the predicted viewport, and then dynamically adjusts the bitrate adaptation strategy with QoE as the optimization objective. 
As MFVP reports the probability for each tile being watched by the user, we set $p_{vp}$ as the threshold for classifying the tiles to be in or out of the viewport. 
We use $Dis(i,j)$ to represent the minimum Manhattan distance between tile $(i,j)$ and all tiles $(u,v)$ with $p_{u,v}\in[p_{vp},1]$. 
Due to the wrapping around property, the maximum possible distance between two tiles is $m+n/2$. 
We divide tiles into $k$ classes, and $k$ should be less than or equal to the number of selectable bitrate levels $l$, here we let $k=l$. The class of tile $(i,j)$ is $rank(i,j)=max(k-Dis(i,j),1)$. 
Fig.~\ref{ba} demonstrates the classification results using 3 test videos from the dataset~\cite{lo2017360}. 
The objective of our bitrate adaptation problem evaluates the user QoE from two aspects. The first QoE metric corresponds to the basic perceived video quality, which is contributed by all the tiles in a chunk $c$: $Q_1^c=\text{Confidence}(SF) \frac{1}{mn} \sum_{i=1}^{m} \sum_{j=1}^{n}rank(i,j) x_{i,j}$, where $x_{i,j}$ is the bitrate assigned to tile $(i,j)$. Since changing $SF$ will affect the prediction accuracy, we use the function $\text{Confidence}(SF) \in (0,1)$ to define the confidence level of MFVP prediction results. The second QoE metric is the quality change which consists of two parts: the inter-chunk change and the intra-chunk change. The former captures the quality change between neighboring chunks. It is defined as the average change of the average consumed tiles’ qualities between consecutive chunks. The intra-chunk change quantifies the variation of qualities of consumed tiles belonging to the same chunk. Specifically, we define the quality change as $Q_2^c=|Q_1^c-Q_1^{c-1}|+StdDev\{x_{i,j}| rank(i,j)>1\}$. Our optimization problem is to find appropriate bitrates $b_i$s that maximize the user QoE, which can be formulated as follows.

\begin{align}	
	\max \text{	}	& Q_1^c-\lambda Q_2^c \label{eq:obj}\\
	\textrm{s.t.	} & \sum_{i=1}^{m} \sum_{j=1}^{n} x_{i,j} + B_{SM}(m*n) \leq B_c  \label{eq:c1}\\
	& T_{predict}^c(SF) < L \label{eq:c2}\\
	& x_{i,j} \in [R_L, R_H], \forall i\in [1,m], \forall j\in [1,n] \label{eq:c3}
\end{align}


Eq.~\ref{eq:obj} is the objective and the other equations present the constraints. As shown in Eq.~\ref{eq:obj}, the overall objective is the weighted sum of the described two QoE metrics, where $\lambda$ is the weight parameter reflecting the application preference. The first constraint is the bandwidth constraint in Eq.~\ref{eq:c1}. We use $B_c$ to denote the available bandwidth for chunk $c$, which can be estimated based on the recent throughput for downloading the last several chunks. 
The transmission cost mainly comes from two parts, i.e., the tiles and the saliency maps. The bandwidth cost for downloading the tiles is $\sum_{i=1}^{m} \sum_{j=1}^{n} x_{i,j}$.
$B_{SM}(m*n)$ denotes the bandwidth cost for transmitting the saliency maps and it is affected by the number of images and the resolution of each saliency map. 
The former is determined by the transmission rate of the saliency maps which we set to one saliency map per chunk, while the latter depends on the downsampling ratio, which is set to 144.
Then the bandwidth consumption for the saliency maps is directly affected by the resolution after downsampling.
This constraint implies that we always try to avoid rebufferring and maintain a smooth playback.
Eq.~\ref{eq:c2} indicates the second constraint $L$ that the prediction must be made on time to meet the stringent delay requirement in live streaming. The time for predicting the tiles in the viewport for a future chunk, i.e., $T_{predict}^c(SF)$, should be less than the chunk length, so that the prediction results can be available when making bitrate decisions. 
Finally, in Eq.~\ref{eq:c3}, we restrict the bitrate selection within a range between $R_L$ and $R_H$, which denote the lower and the upper bounds of the assigned bitrates.

%
%
%
%
%
%

\subsubsection{Efficient Solution}
To solve the formulated problem, we first specify the selections of key parameters such as 
$SF$. 
We set $SF$ as a highest possible value to keep $T_{predict}^c(SF)$ satisfying Eq.~\ref{eq:c2}. 
Then we model $\text{Confidence}(\cdot)$ as a sigmoid-like function of $SF$ as shown in Fig.~\ref{fig:sf-f1}. 
After deciding $SF$, 
the QoE optimization can be solved through exhaustive search because of the small instance size. Since the optimization should be invoked at a high frequency, it is still challenging due to the large search space. To support real-time optimization, we need to efficiently prune the search space. To this end, we identify the following important constraints and opportunities for boosting the solution’s efficiency: (1) all tiles belonging to the same class should have the same quality level. This constraint allows us to perform bitrate adaptation on a per-class basis instead of on a per-tile basis, which significantly reduces the search space, (2) the bitrate of a tile should not be lower than that of any other tile in the same chunk with a lower class, and (3) the bitrate selection should consider the constraint of throughput, i.e., Eq.~\ref{eq:c1}.


After the bitrate decision is made, the client makes appropriate requests for tiles of different resolutions to the request handler on the server, after which the requested tiles for the future chunks are transmitted, buffered, and then played in the client's player.

\begin{figure*}
	\centering
	\begin{subfigure}{0.32\textwidth}
		\includegraphics[width = 1\linewidth]{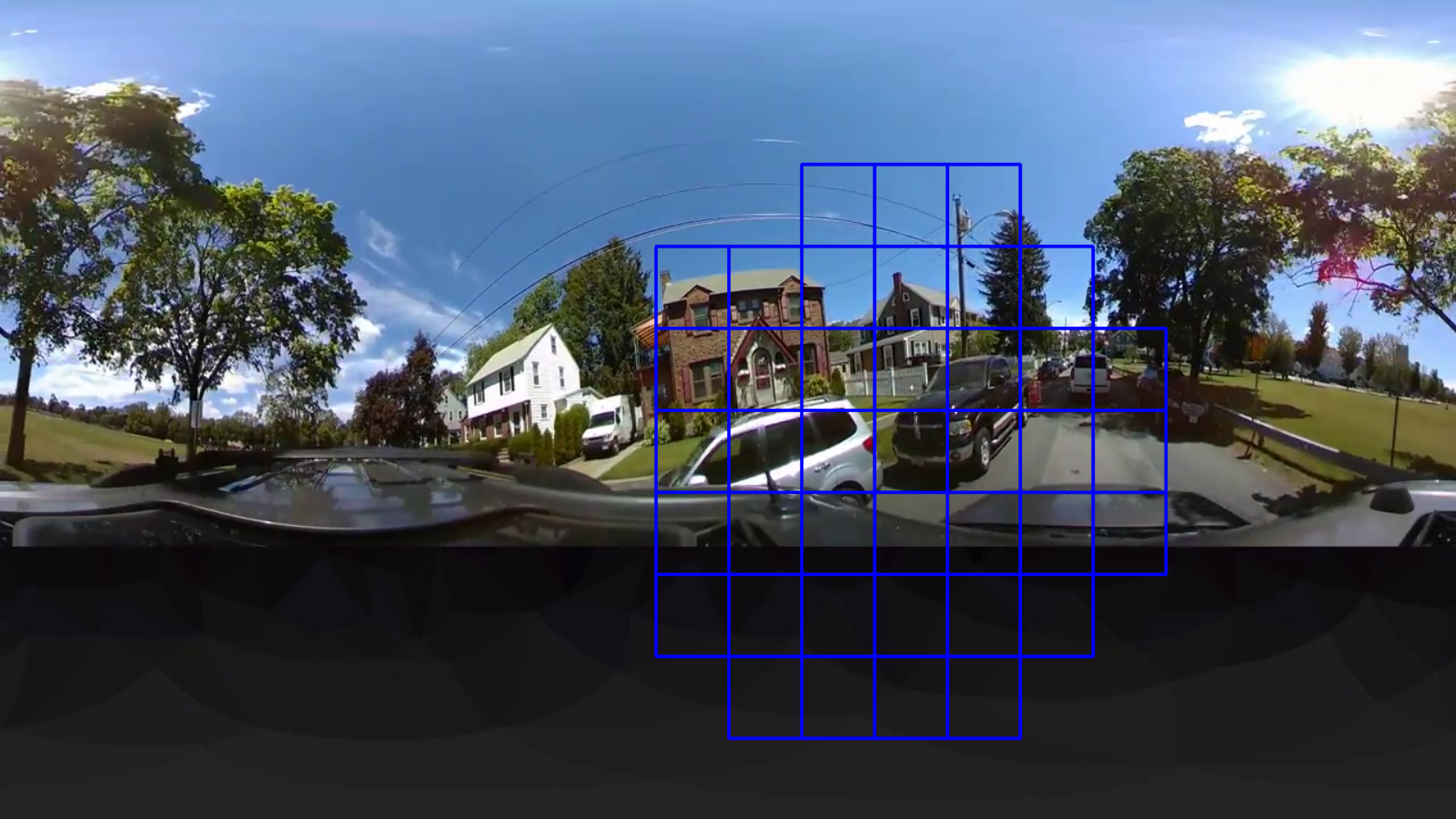}
		\caption{\small drive}
		\label{infer_drive}
	\end{subfigure}	
	\hfil
	\begin{subfigure}{0.32\textwidth}
		\includegraphics[width = 1\linewidth]{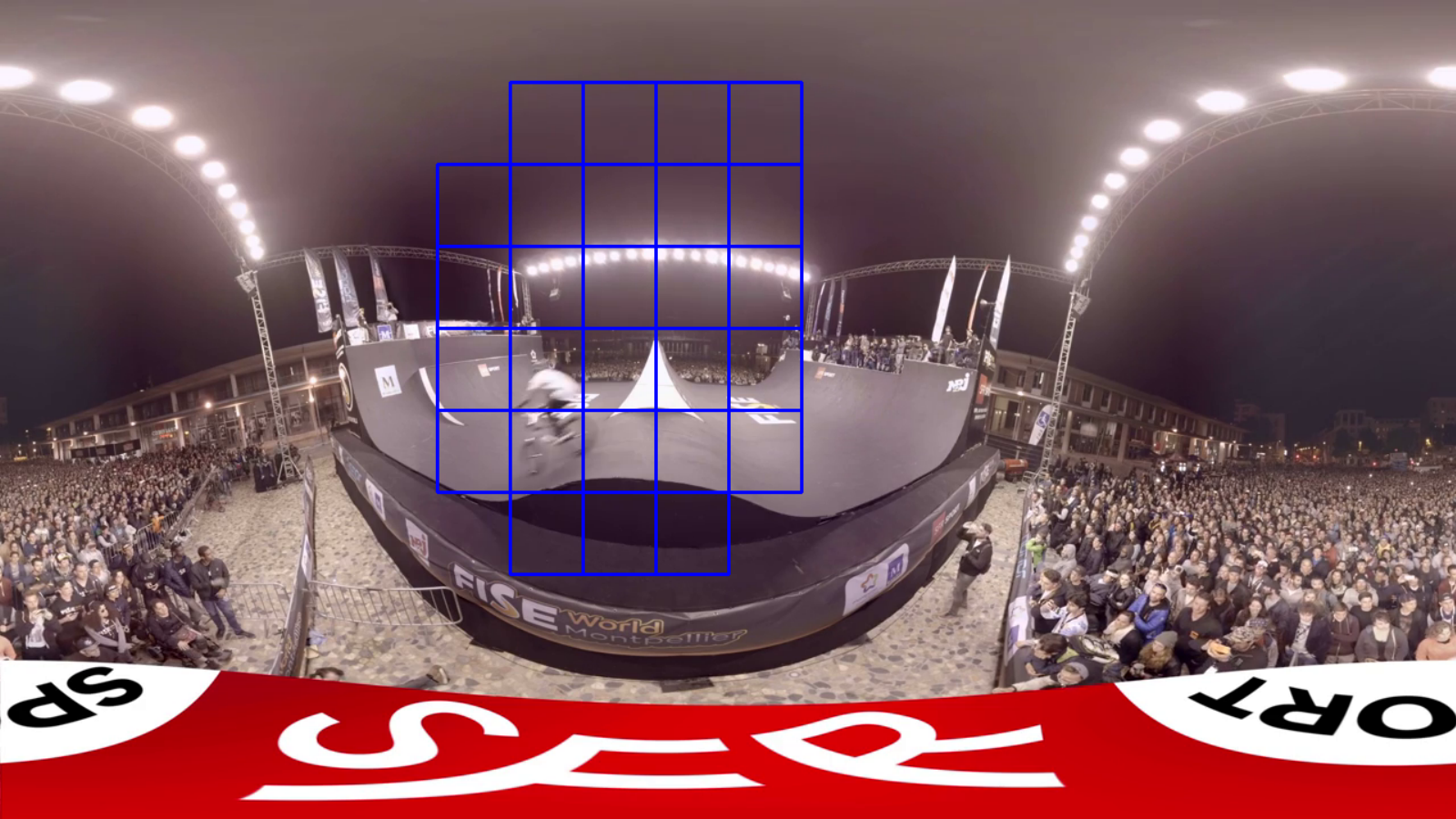}
		\caption{\small sport}
		\label{infer_sport}
	\end{subfigure}	
	\hfil
	\begin{subfigure}{0.32\textwidth}
		\includegraphics[width = 1\linewidth]{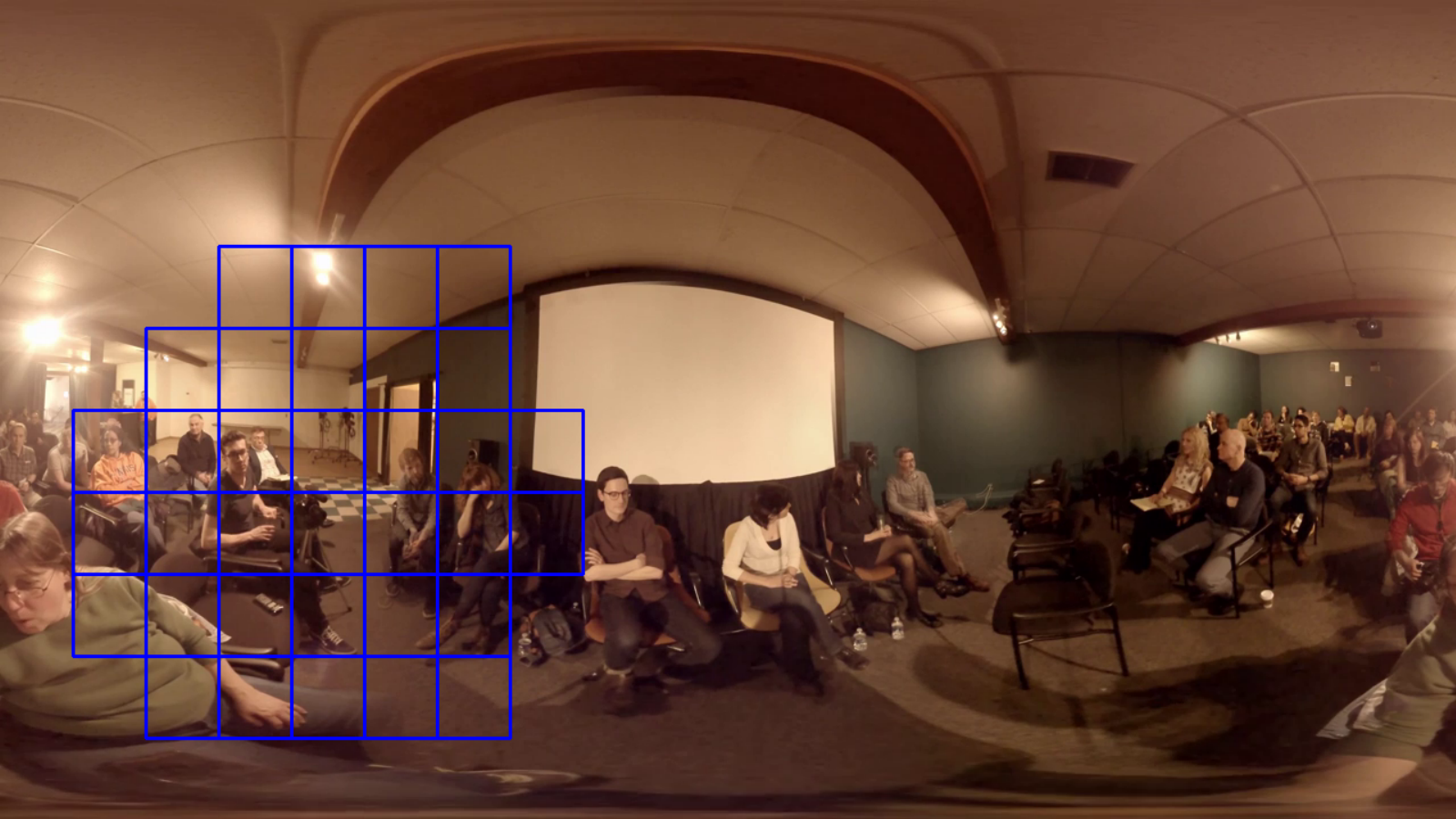}
		\caption{\small panel}
		\label{infer_panel}
	\end{subfigure}

	\caption{Snapshots of viewport prediction results achieved by MFVP using 3 test videos from~\cite{lo2017360}. The blue rectangles indicate the viewports predicted by MFVP.}
	\label{infer}
\end{figure*}

\begin{figure*}
	\centering
	
	\begin{subfigure}{0.32\textwidth}
		\includegraphics[width = 1\linewidth]{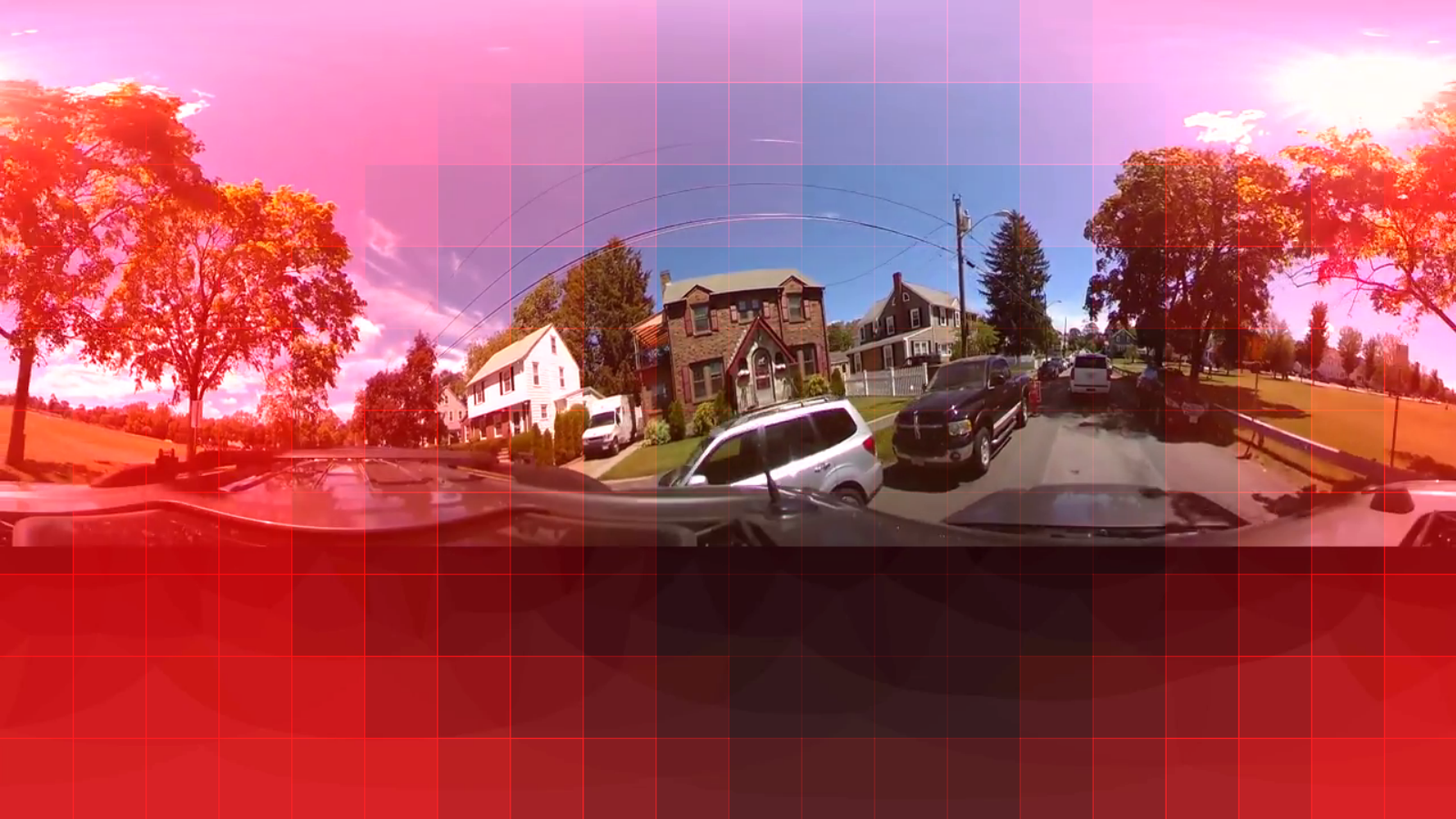}
		\caption{\small drive}
		\label{ba_drive}
	\end{subfigure}	
	\hfil
	\begin{subfigure}{0.32\textwidth}
		\includegraphics[width = 1\linewidth]{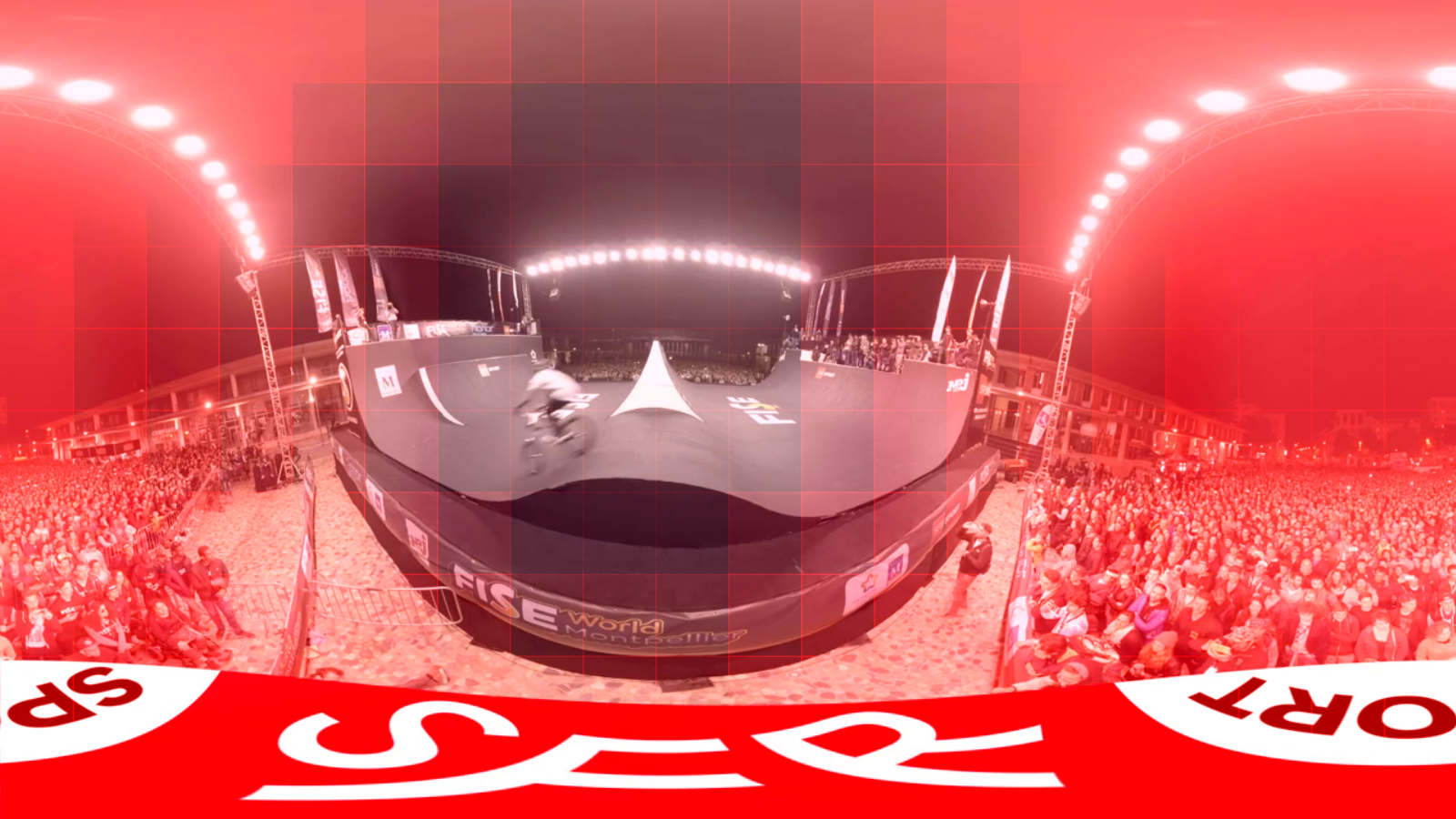}
		\caption{\small sport}
		\label{ba_sport}
	\end{subfigure}	
	\hfil
	\begin{subfigure}{0.32\textwidth}
		\includegraphics[width = 1\linewidth]{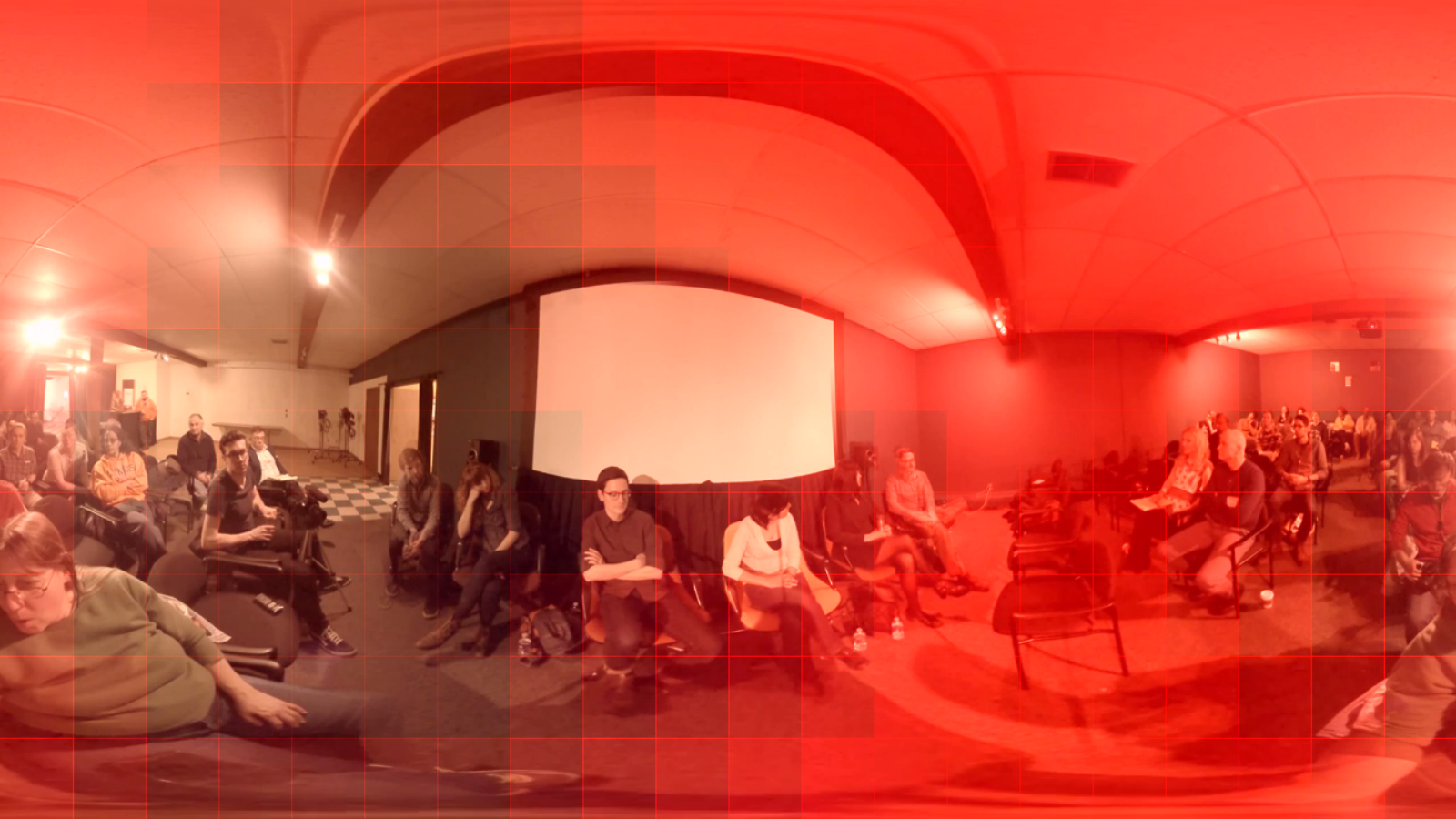}
		\caption{\small panel}
		\label{ba_panel}
	\end{subfigure}

	\caption{CBA classifies tiles using the viewport prediction results shown in Figure~\ref{infer}. 
		The figures depict the classification of tiles with 10 × 20 tiled frames.
		Tiles with higher transparency have a higher class and may be assigned a higher bitrate.
		Tiles with the same transparency belong to the same class.
	}
	\label{ba}
\end{figure*}

\begin{figure}
	\centering
	\includegraphics[width=0.7\linewidth]{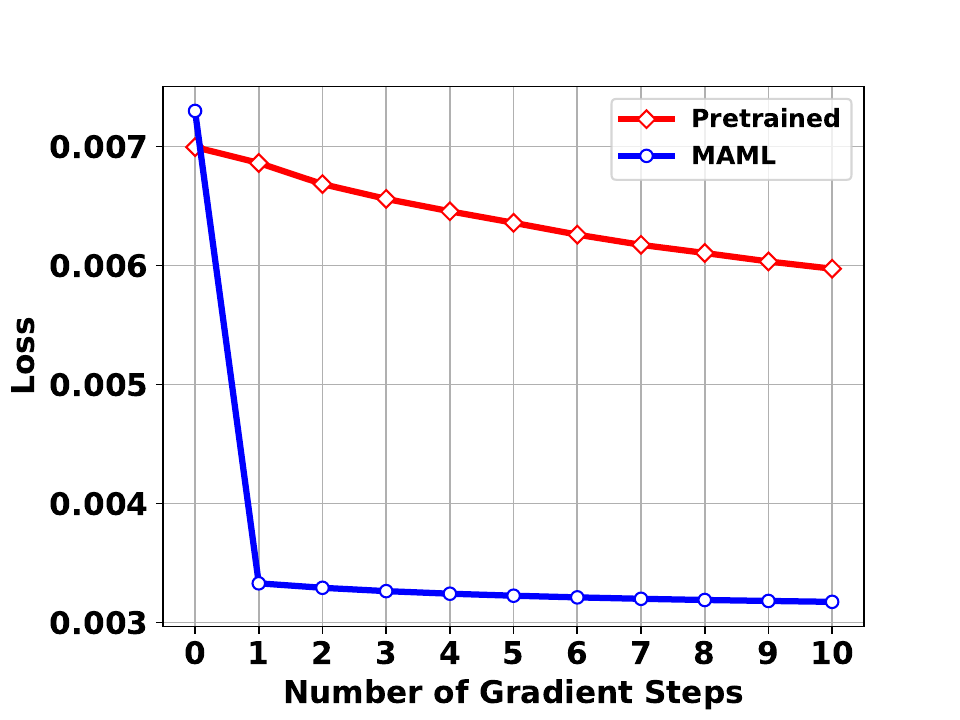}
	\caption{Learning Curve of Fine-tuning.}
	\label{fig:MAML_result}
\end{figure}

\section{Performance Evaluation}
\label{sec:experiment}
In this section, we conduct extensive experiments to evaluate the prediction performance of MFVP and the streaming performance of our bitrate adaptation algorithm. 

\subsection{Methodology}
\subsubsection{Data Traces}
The data traces used in our evaluation are collected from our real-world measurements and some other open data sets. \textcolor{black}{The reason we choose this bandwidth is because he is a real bandwidth data, which can reflect the effect of the algorithm running in the actual system. The characteristics of this data set and the reasons why we chose it are as follows. (1) The data is collected in the 4G network in Ghent, Belgium, and the research focuses on the 4G/LTE network, which is consistent with our mobile scenario. Actual measurements: The dataset was constructed over multiple routes, measuring the available bandwidth when downloading large files over HTTP. (2) Proper download speeds are guaranteed using dedicated servers in iLab.t’s virtual wall infrastructure connected to 100 Mb/s Ethernet. It can meet some of the bandwidth requirements of 360 video scenes. (3) Data recording was performed using an Android application running on a smartphone (Huawei P8 Lite) connected via 4G. The measurement device is also a mobile device. (4) Complete statistical information, recording various attributes including GPS coordinates, number of bytes received since the last data point, number of milliseconds since the last data point, etc. Through the last two pieces of data, the average throughput can be obtained. (5) Different modes of transportation were measured and the throughput logs of six modes of transportation were collected: walking, bicycle, bus, tram, train and car. (6) There are link constrained scenarios and it is observed that when connectivity is constrained (e.g. due to tunnels, large buildings and general poor coverage), the throughput values are lower. The type of transport and the route chosen have a significant impact on the available bandwidth. It is also the bandwidth fluctuation required for our experiments.}
To the dynamic network conditions, we replay the bandwidth traces from a 4G/LTE dataset captured during mobility~\cite{van2016http}. \textcolor{black}{Our main contribution is to analyze the main overhead and cost of complex mobile prediction and adaptive streaming tasks, carefully split the computation on both sides of the flow, and reduce the size of the saliency map by reducing the data sampling frequency/transmission rate and compression, learning model to minimize its communication overhead. The code rate adaptation model is not the focus of our consideration. Therefore, our code rate adaptation model is relatively simple and the effect is not obvious when the bandwidth is too fluctuating. The bandwidth is scaled to a reasonable range. In the scaled bandwidth , it also has certain bandwidth changes, reflecting the effect of our design.} 

We train and validate our MAML-based few-shot fast trainer on Salient360Video dataset~\cite{david2018dataset}. 
It contains 19 360\textdegree~videos and saliency maps corresponding to video frames.
The first 360\textdegree~video viewing traces are from the open dataset~\cite{lo2017360}, which includes ten videos freely viewed by 50 users each with each video watched for 60 seconds. We randomly divide the 50 users into two subsets: 80$\%$ for training and 20$\%$ for testing. \textcolor{black}{The second 360\textdegree~video viewing traces are from the open dataset~\cite{david2018dataset}, which includes nineteen videos freely viewed by 57 users each with each video watched for 20 seconds. Similar, we randomly divide the 57 users into two subsets: 80$\%$ for training and 20$\%$ for testing. } We use the traces from the train set to train the proposed network. 
We replay all the 360\textdegree~video watching events in the test set for testing and for each video watching event randomly select 10 bandwidth traces from the 4G/LTE dataset.

\subsubsection{Algorithms for Comparison}
We compare MFVP with the representative existing viewport prediction algorithms, which are Linear Regression (LR)~\cite{qian2018flare}, SalGCN~\cite{lv2020salgcn}, LiveDeep~\cite{feng2020livedeep}, \textcolor{black}{AME~\cite{li2019very,sun2022live}}, and MobileNetv2(MN)~\cite{sandler2018mobilenetv2}. LR treats the viewport trajectory as time series and estimates the future viewport using a linear model; SalGCN predicts saliency maps using graph convolutional networks, then divides the saliency maps into multiple regions according to the number of tiles, calculates the average score of each region, 
and uses the high score area as the predicted viewport;
LiveDeep uses a neural network model that mixes CNN and LSTM to predict the viewport; MN replaces the CNN network of LiveDeep with MobileNetv2 which is suitable for mobile terminals, and the other parts remain unchanged. \textcolor{black}{Since Li et al.~\cite{li2019very} and Sun et al.~\cite{sun2022live} technical routes of these two jobs are relatively consistent, we select AME as our baseline to represent this type of work. For the reproduction of AME, we choose Utilizing saliency maps derived from video sequences. Because the performance of using other user heat maps is not much different from that of using Utilizing saliency maps derived from video sequences. And such an approach is more conducive to user privacy protection and a more realistic approach.}
For the rate adaptation, we compare our bitrate adaptation algorithm with the pyramid-based bitrate allocation scheme(PBA) used in PARIMA~\cite{chopra2021parima}, which chooses qualities intelligently with a gradually decreasing quality according to the distance from the predicted viewport.

\subsubsection{Settings and Metrics}
As the functions such as $B_{SM}(m*n)$ and $T_{predict}^c(SF)$ can be mostly dependent on the image size of saliency map or the client's hardware, we can fit them by using the measurement data prior to the video streaming.
We use 10x20 as the tiling setting. 
For the weights in the QoE objective, we set $\lambda=2$. 
For throughput prediction, we adopt a moving average predictor~\cite{yao2008empirical} based on the past 5 samples.
We consider six quality levels with different bitrate settings: (1) 360p (1Mbps), (2) 480p (2.5Mbps), (3) 720p (5Mbps), (4) 1080p (8Mbps), (5) 2K (16Mbps), and (6) 4K (40Mbps). To evaluate the performance, we examine accuracy, running time, average quality level, quality level change, used bandwidth and rebuffering time during the video playback.
Note that because the video data are requested chunk by chunk in a typical video streaming system, the union of tiles viewed in a chunk is taken as ground truth. 

\subsection{Result and Analysis}

\subsubsection{Performance of MAML-based Fast Trainer for Saliency Prediction Networks}
We train and test the effectiveness of our MAML-based fast trainer in the Salient360Video dataset~\cite{david2018dataset}. 
We use 80$\%$ prediction tasks for the meta-learning training and others for the test. 
Each task contains 20 samples randomly drawn from one video. 
For each task, the support set consists of 5 samples and the query set consists of the other 15 samples.
We fine-tune the initial model on the support set for each task and evaluate it on the query set. 
We compare the performance of our trainer on new videos with the pretrained model using the same training task data.
As shown in Fig.~\ref{fig:MAML_result} , using only 5 samples from a new video prediction task and 10 epochs of update, MAML consistently obtains a network that performs well on the new task. 
The model learned using MAML rapidly improved in performance after one gradient step and consistently outperformed the pretrained model significantly in subsequent gradient steps. 
This suggests that MAML optimizes the parameters so that they are in a region that is easy to adapt quickly and is sensitive to loss functions from $p(T)$.
In conclusion, our trainer trains quickly in live scenarios, which can help us achieve great real-time performance.

\begin{figure}
	\centering
	\includegraphics[width=0.98\linewidth]{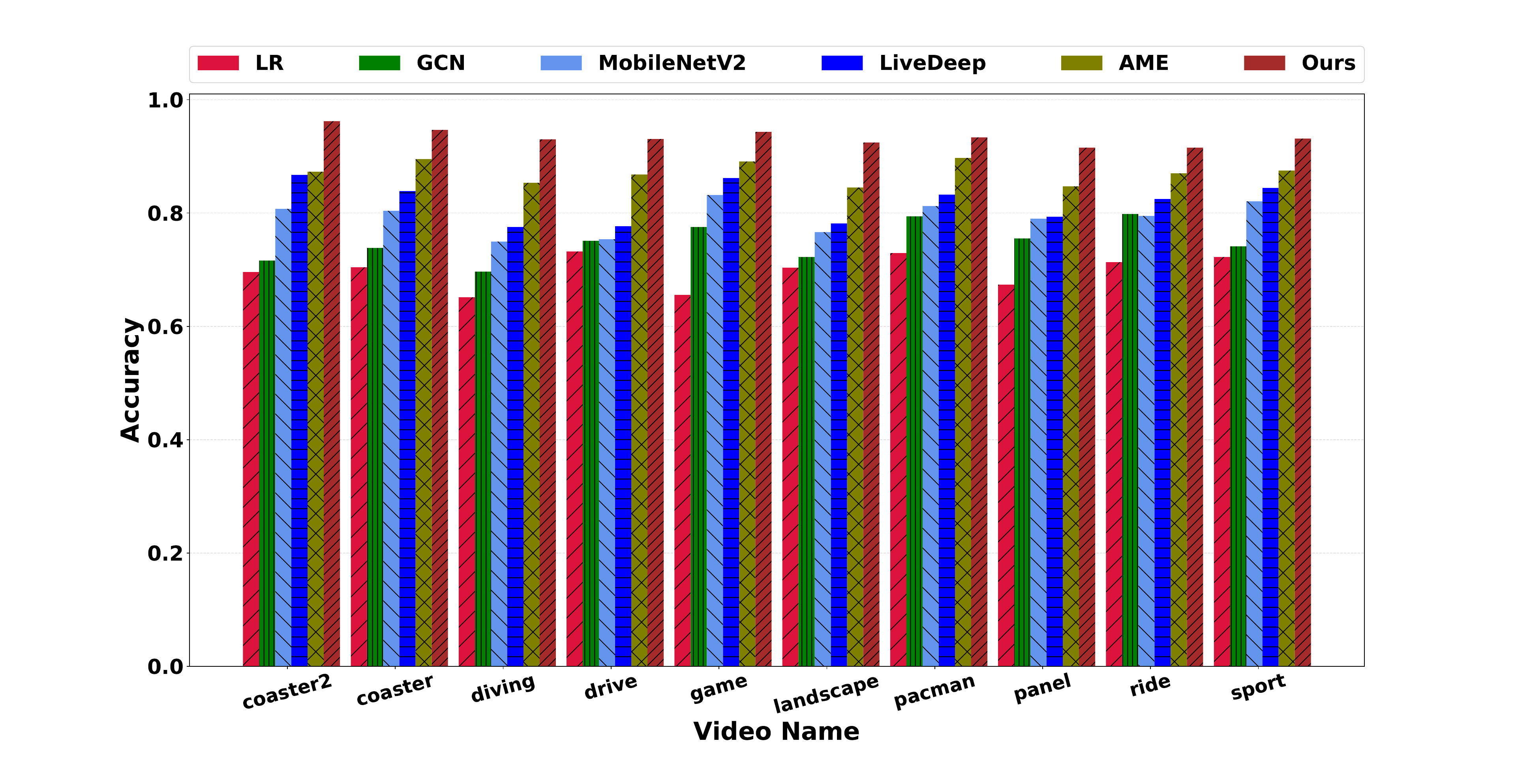}
	\caption{ Overall prediction accuracy with the comparison among the six methods on dataset 1.}
	\label{fig:acc_1}
\end{figure}

\begin{figure}
	\centering
	\includegraphics[width=0.98\linewidth]{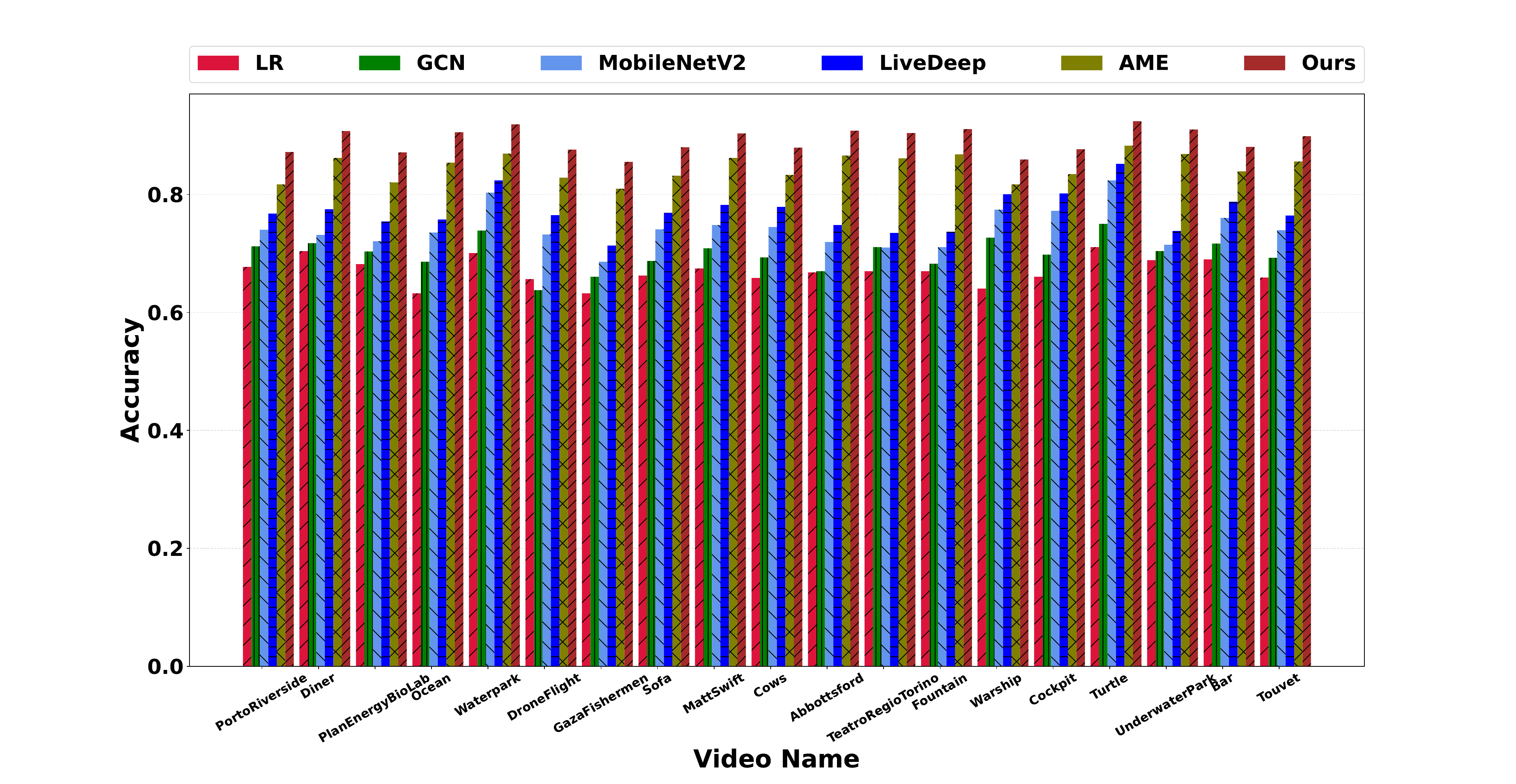}
	\caption{ \textcolor{black}{Overall prediction accuracy with the comparison among the six methods on dataset 2.}}
	\label{fig:acc_2}
\end{figure}

\subsubsection{Prediction Performance of MFVP}
\textcolor{black}{ We first examine the performance gain from MFVP approach in terms of the prediction accuracy. 
Fig.~\ref{fig:acc_1} \textcolor{black}{and Fig.~\ref{fig:acc_2}} show the prediction accuracy of MFVP and other comparison algorithms for the two dataset when the prediction window is 1 second, which shows the clear performance improvement. 
We observe that MFVP achieved high prediction accuracy (i.e., $>90\%$) over the 10 test videos for dataset 1, 
and the accuracy of MFVP is 8.1$\%$ to 28.7$\%$ higher than other algorithms. We observe that MFVP achieved high prediction accuracy (i.e., $>90\%$) over the 10 test videos for dataset 2, 
and the accuracy of MFVP is 4.7$\%$ to 6.7$\%$ higher than other algorithms.
Since the viewport prediction is reported chunk by chunk, the number of tiles in the viewport of different chunks varies widely, and it is difficult for LR to capture this change. SalGCN does not consider user history information. LiveDeep and MN only use video frames and single-user history information and do not use saliency maps to consider the interests of multiple users. AME uses video frames and corresponding saliency maps for prediction.}

MFVP approach can better predict user behavior by utilizing both of the spatial information of cross-user interest distribution and the temporal information of historical user viewports.

Figure~\ref{infer} demonstrates the viewport prediction results using 3 test videos from the dataset~\cite{lo2017360}. 
The blue rectangles represent the viewports predicted by MFVP. 
Figure~\ref{infer_drive} shows a video captured by a moving camera, where the surrounding environment moves with the camera, creating a challenging case for viewport prediction algorithms. 
Figure~\ref{infer_sport} shows a video containing a single attractive region captured by a single camera. In this case, the target area appears to be fixed at the center and the background is static, making it a relatively simple case for viewport prediction. 
Figure~\ref{infer_panel} shows a video with multiple attractive visual objects, which is relatively difficult for viewport prediction due to possible view switching.

\subsubsection{Ablation Study}
In this section, we analyze the main components of modified ConvLSTM which proposed in section 3.3 through ablation experiments. 
Tab.~\ref{tab:ablation} shows ablation experiments’ results respectively, 
where D.S.C. stand for depthwise separable convolution. 
The bold number represents the best performance, and the underlined number represents the second-best performance.
Comparing the line(a) and the line(b), the depthwise separable convolution saves 11$\%$ of inference time due to fewer convolution operations, but relatively, the prediction performance drops a bit.
Comparing the line(a) and line (c), we note that the prediction performance has improved,  because the SE block allows the model to pay more attention to the most informative channel features and suppress the unimportant channel features, but this will brings some extra overhead. 
Line(d) shows that the cooperation of the two operations both improves prediction performance and reduces inference time. 
These ablation experiments illustrate that the effectiveness of using depthwise separable convolution and SE block in ConvLSTM for viewport prediction.

\begin{table}[]
	
	\renewcommand{\arraystretch}{1.3}
	\caption{Performance of different component of modified ConvLSTM for viewport prediction. 
	}
	\label{tab:ablation}
	\centering
	\scalebox{0.9}
	{
		\begin{tabular}{c|cc|ccc}
			\hline
			Methods & D.S.C.                     & SE                        & Accuracy(\%)   & F1-Score(\%)   & \begin{tabular}[c]{@{}c@{}}Inferring\\ Time(ms)\end{tabular} \\ \hline
			(a)     & ×                         & ×                         & 93.29          & 85.57          & 350                                                          \\
			(b)     & \checkmark & ×                         & 93.02          & 85.41          & \textbf{309}                                                 \\
			(c)     & ×                         & \checkmark & \textbf{93.81} & \textbf{86.75} & 372                                                          \\
			(d)     & \checkmark & \checkmark & \underline{93.71}          & \underline{86.60}          & \underline{330}                                                          \\ \hline
		\end{tabular}
	}
	\vspace{0.1cm}
\end{table}

%

\begin{table}[]
	\begin{center}
		\renewcommand{\arraystretch}{1.3}
		\caption{Comparison of different models in FLOPs, model file size, and inferring time. 
		}
		\label{tab:mobile}
		\centering
		\scalebox{1.1}
		{
			\begin{tabular}{c|ccc}
				\hline
				Model                                                     & FLOPs         & Model Size & Inferring Time \\ \hline
				LR                                                          & \textless{}1M & \textless{}1KB        & \textless{}1ms \\
				MN                                                          & 1401M         & 13790KB    & 483ms          \\
				ConvLSTM & 149M          & 710KB      & 330ms          \\ \hline
			\end{tabular}
		}
	\end{center}
	\vspace{0.1cm}
\end{table}

\subsubsection{Feasibility of MFVP on Mobile Terminal}
We next check the model complexity to verify the feasibility of applying MFVP on mobile clients.
Since MFVP places the saliency prediction on the server and it requires no engagement from the user during the video watch event, we only need to consider the complexity of the viewport prediction model.
To measure the computation complexity, a widely used metric is FLOPs, i.e. the number of floating-point multiplication-adds. However, FLOPs is an indirect metric. It is an approximation of, but usually not equivalent to the direct metric that we really care about, such as speed or latency~\cite{Ma_2018_ECCV}. 
Therefore, using FLOPs as the only metric for computation complexity is insufficient. 
So we also compare inferring time on mobile devices and TensorFlow Lite model file size.
We do not report SalGCN numbers because graph convolutions are not yet supported on mobile.
We also do not report LiveDeep numbers because its model is too large and not practical for mobile devices.
The results are shown in Tab.~\ref{tab:mobile}.
In the last column we report inferring time in milliseconds (ms) for Huawei Mate 30 (using TensorFlow Lite).
Although LR runs fast, its accuracy is low and only works for short prediction windows (see Fig.~\ref{fig:acc_1}).
Our modified ConvLSTM model is not only fast enough(i.e., inferring time is less than a chunk length), but also the most accurate of the three models.
Notably, our modified ConvLSTM model is 1.4× more faster and 19× smaller while still outperforms MN. 
Furthermore, we observe that although the FLOPs of our modified ConvLSTM model is only one tenth of that of MN, the inferring time is not one tenth of that of MN. 
This is because the parallelism of MN is higher, while the parallelism of ConvLSTM is lower since the calculation of ConvLSTM at the current moment depends on the result of the previous moment.


%

\begin{figure}
	\centering
	\begin{minipage}{0.49\linewidth}
		\centering
		\includegraphics[width=0.95\linewidth]{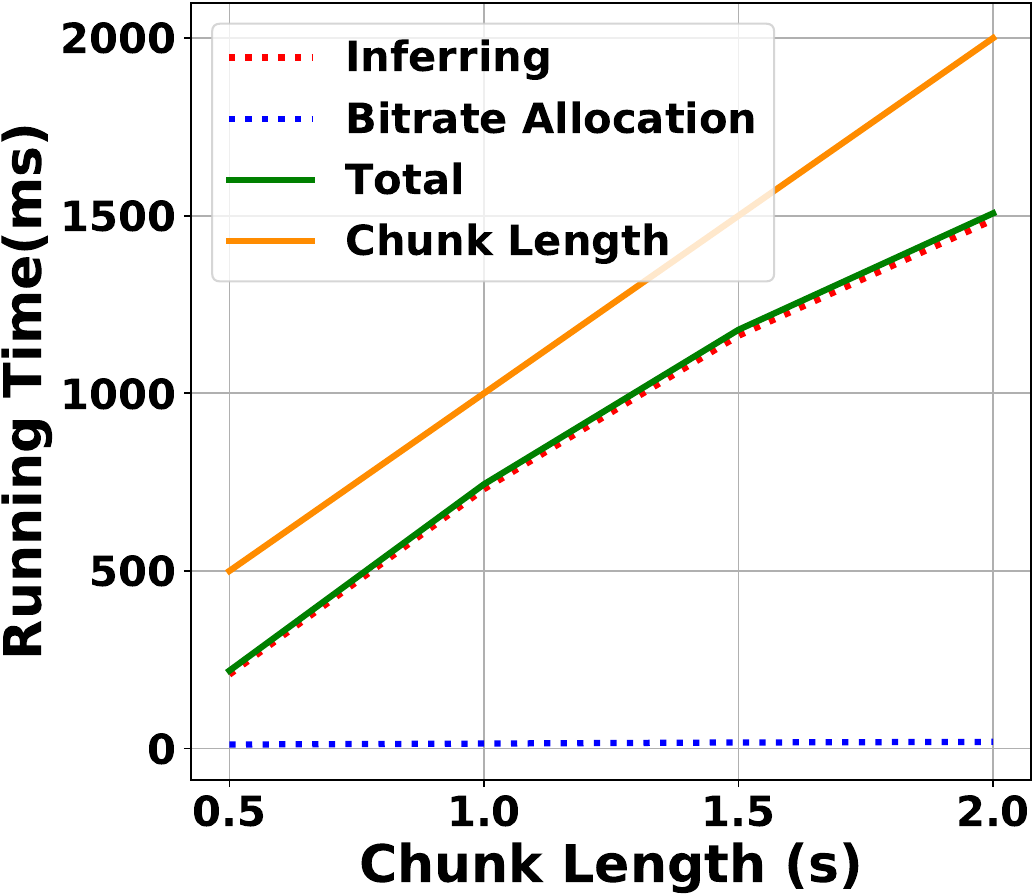}
		\caption{\small Comparison of chunksize with inferring time and bitrate adaptation time}
		\label{time1}
	\end{minipage}
	\begin{minipage}{0.49\linewidth}
		\centering
		\includegraphics[width=0.95\linewidth]{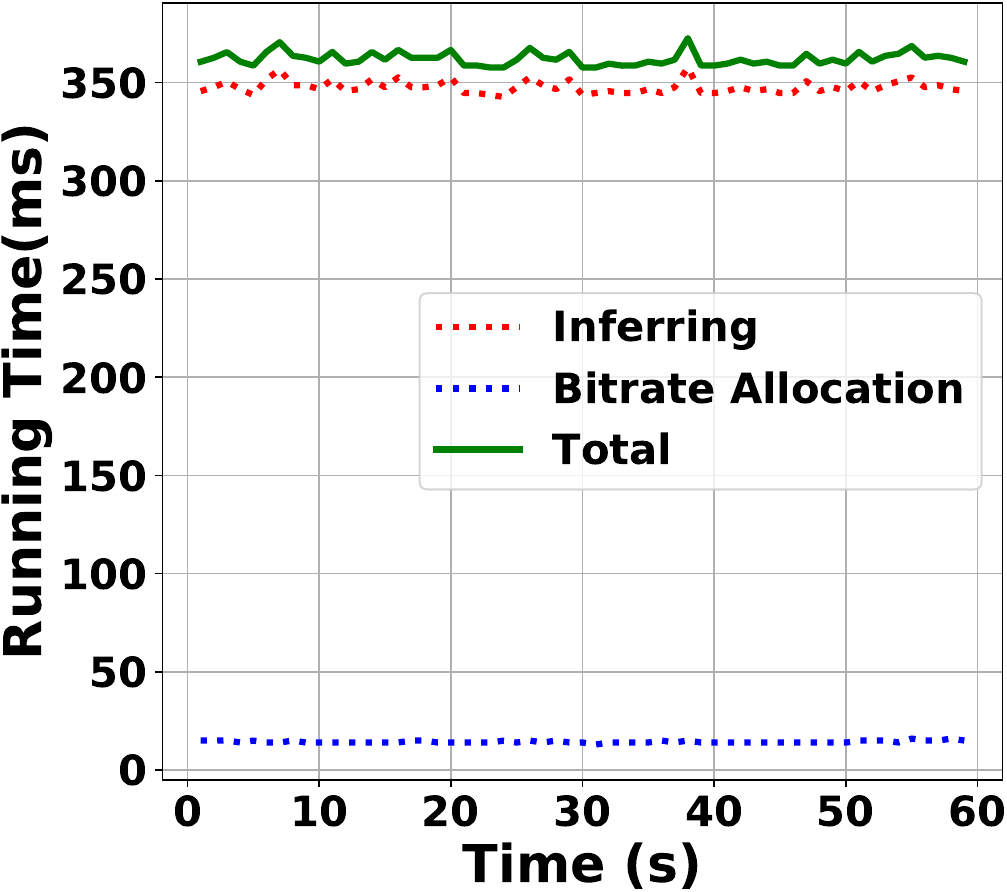}
		\caption{\small Inferring time and bitrate adaptation time in a watch event}
		\label{time2}
	\end{minipage}
\end{figure}

\begin{figure*}
	\centering
	\includegraphics[width=0.95\linewidth]{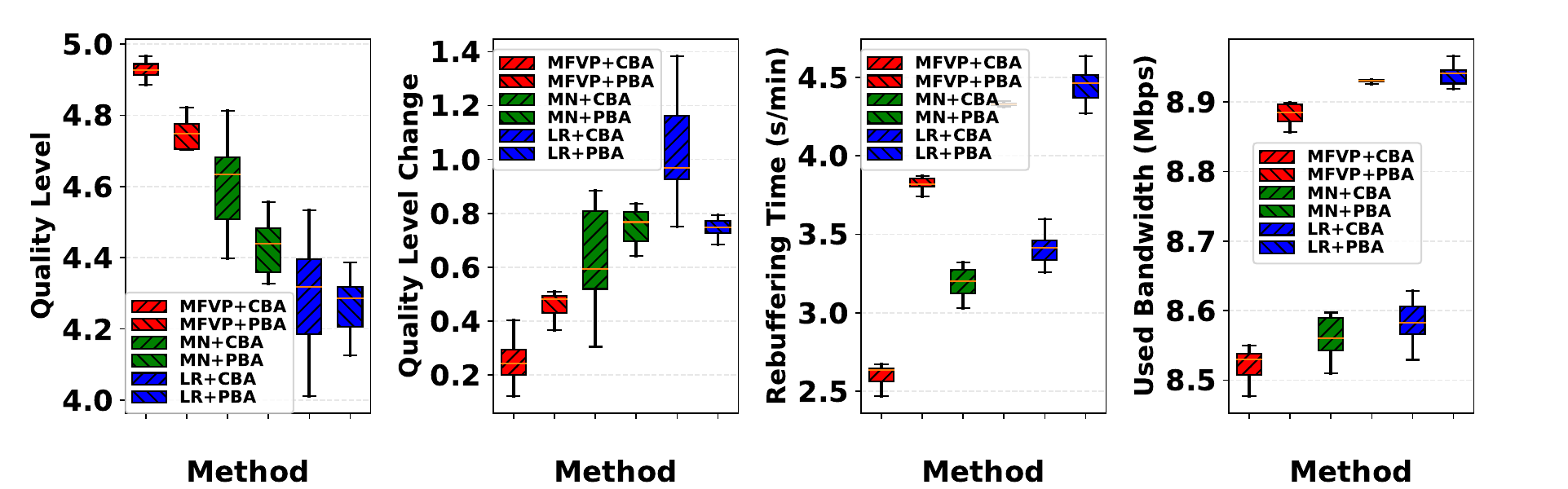}
	\caption{Comparison of different algorithms in average quality level, quality level change, rebuffering time and bandwidth consumptions}
	\label{fig:qoe}
\end{figure*}

\subsubsection{Feasibility of MFVP in Live Streaming}
We next check the prediction timeliness 
and communication overhead
to verify the feasibility of applying MFVP in a live streaming scenario for mobile clients. In order to ensure a smooth viewing experience, the processing time of each video segment must be less than the segment duration. Otherwise, users may experience rebuffering. In our case, the processing time is the sum of the inferring time and the bitrate adaptation decision time. 
Fig.~\ref{time1} shows that for different chunk lengths, the average processing time on Huawei Mate 30 is always less than the corresponding chunk length. 
This is because our method can always select the appropriate parameter $SF$ according to different conditions (e.g., mobile phone performance and chunk length).
Another observation is that the growth rate of processing time is less than the growth rate of chunksize, which means that even with larger chunksize, we can still meet real-time requirements. We further check a representative video watching event (User 47 watching Video 'coaster') in Fig.~\ref{time2}. We can clearly observe that the running time of MFVP and CCPA is relatively stable, which means that there will be no accidental rebuffering events due to running time jitter.


\begin{figure}
	\centering
	\includegraphics[width=0.75\linewidth]{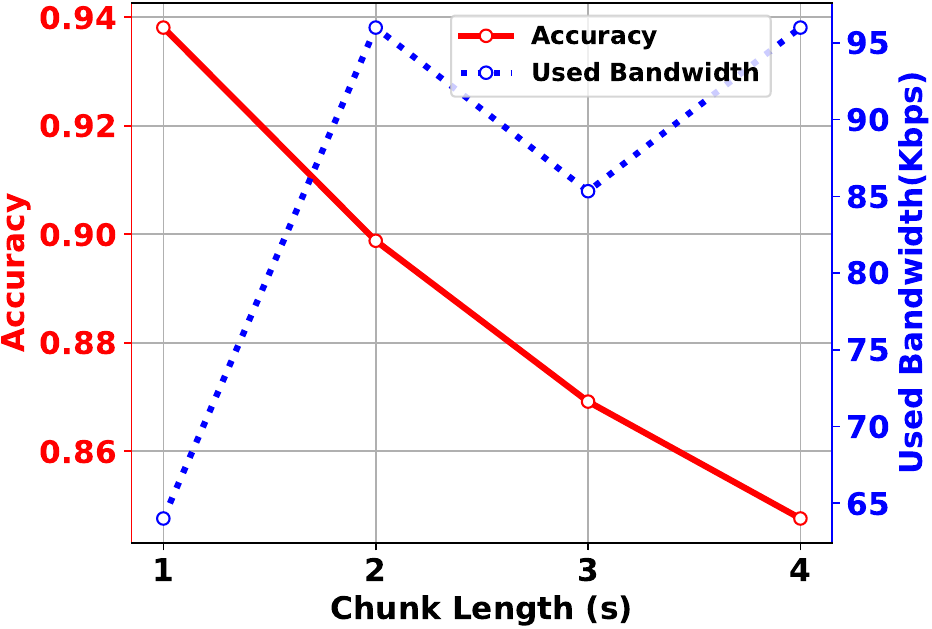}
	\caption{Impact of chunk length on the transmission overhead of saliency maps and prediction performance.}
	\label{fig:chunk}
	\vspace{-0.3 cm}
\end{figure}

\subsubsection{Trade-off Between Prediction Performance and Feasibility} 
In a live 360-degree video streaming system, the 360-degree video is divided into small segments according to the Dynamic Adaptive Streaming over HTTP(DASH) standard, and each video chunk is usually 1 to 4 seconds long~\cite{bouzakaria2014overhead}.
In order to ensure the feasibility of the prediction algorithm (i.e., reasonable prediction time and as little communication overhead as possible), we need to choose an appropriate video chunk length, 
because it will affect the number of saliency predictions, sampling frequency and downsampling ratio, while these in turn affect viewport prediction performance. Therefore, it is necessary to analyze the interaction of these parameters to 
trade off the viewport prediction performance and its feasibility.

The prediction timeliness is mainly affected by the sampling frequency and the downsampling ratio.
According to Fig.~\ref{dr-time} and~\ref{fig:ratio-f1}, the downsampling ratio above a certain level brings a limited increase in prediction performance, but the prediction timeliness increases linearly.
In our experiments, we find that empirically setting downsampling ratio to 144 can usually serve the purpose, shown as the dashed line in Fig.~\ref{fig:ratio-f1}.
To ensure a smooth viewing experience, we set a sampling frequency that makes the prediction timeliness less than the chunk length.
Otherwise, users may experience rebuffering.
For the communication overhead introduced by MFVP, it consists of two parts, which are the one-time transmission of the ConvLSTM model and the continuous transmission of saliency maps during the streaming.
Because the quantized and compressed model is small in size and only needs to be transmitted once, its transmission overhead is negligible.
Since the saliency prediction on the server side takes some time (about 0.65 seconds to predict once on Nvidia Tesla V100 GPU), different chunk lengths will affect the number of saliency predictions, which in turn affects the bandwidth consumption of transmitting the saliency maps and subsequent viewport prediction performance.
We explore the effect of chunk length on bandwidth consumption and prediction performance in Fig.~\ref{fig:chunk}, where the blue line represents the bandwidth consumption of the saliency maps. We observe that the bandwidth consumption of transmitting the saliency maps is small because it is downsampled and requires a low transmission rate. In addition, the prediction performance deteriorates as the chunk length becomes longer, because the user behavior is more variable due to the farther future to be predicted and the larger time span.
Therefore, for less transmission overhead and better prediction performance, we set the chunk length to 1 second.

\subsubsection{Streaming QoE of Bitrate Adaptation}
We next check the streaming performance using bandwidth traces on mobile phones. 
Because SalGCN and LiveDeep cannot run on the mobile terminal, we choose MN and LR for comparison in terms of prediction algorithms. 
In addition, the output of PBA is the bitrate instead of the bitrate level, so we map its output to the closest bitrate level. We plot the average quality level, quality level change
rebuffering time, and total bandwidth consumption in Fig.~\ref{fig:qoe}. 
As we can see from the figures, our method significantly outperforms other methods in most cases. 
Our method achieves 3.73$\%$ to 14.96$\%$ higher average quality level and 49.6$\%$ to 74.97$\%$ less quality level change than other algorithms. 
No matter which bitrate adaptation algorithm is used, MFVP will be better than other prediction algorithms because of its high accuracy.
As CBA is designed to work with MFVP, it generally outperforms PBA. The only exception is the quality change under very poor prediction performance (e.g., LR), where we can adjust $\lambda$ to strike a better trade-off between average quality and quality churn.

%
%

\section{Conclusion}
\label{sec:conclusion}

In this paper, we propose an advanced learning-based viewport prediction approach for 360-degree video live streaming and carefully split the computation on both sides of the streaming. 
To meet real-time requirements and better cope with the complexity of live streaming’s content, we use the MAML-based few-shot fast trainer to obtain the saliency prediction network for a new video with only a few samples. Therefore, it can be applied to various video streaming with little extra cost in inference.
We further integrated our viewport prediction with a typical bitrate adaptation scheme for tile-based viewport-adaptive live streaming. The results from the trace-driven simulations demonstrated that our proposed method outperforms other representative state-of-the-art algorithms.


%

%
%
%
%

\ifCLASSOPTIONcaptionsoff
\fi



\bibliographystyle{IEEEtran}
\bibliography{sample-base}

\begin{thebibliography}{10}
\providecommand{\url}[1]{#1}
\csname url@samestyle\endcsname
\providecommand{\newblock}{\relax}
\providecommand{\bibinfo}[2]{#2}
\providecommand{\BIBentrySTDinterwordspacing}{\spaceskip=0pt\relax}
\providecommand{\BIBentryALTinterwordstretchfactor}{4}
\providecommand{\BIBentryALTinterwordspacing}{\spaceskip=\fontdimen2\font plus
\BIBentryALTinterwordstretchfactor\fontdimen3\font minus \fontdimen4\font\relax}
\providecommand{\BIBforeignlanguage}[2]{{%
\expandafter\ifx\csname l@#1\endcsname\relax
\typeout{** WARNING: IEEEtran.bst: No hyphenation pattern has been}%
\typeout{** loaded for the language `#1'. Using the pattern for}%
\typeout{** the default language instead.}%
\else
\language=\csname l@#1\endcsname
\fi
#2}}
\providecommand{\BIBdecl}{\relax}
\BIBdecl

\bibitem{qian2018flare}
F.~Qian, B.~Han, Q.~Xiao, and V.~Gopalakrishnan, ``Flare: Practical viewport-adaptive 360-degree video streaming for mobile devices,'' in \emph{Proceedings of the 24th Annual International Conference on Mobile Computing and Networking}.\hskip 1em plus 0.5em minus 0.4em\relax ACM, 2018, pp. 99--114.

\bibitem{xie2017360probdash}
L.~Xie, Z.~Xu, Y.~Ban, X.~Zhang, and Z.~Guo, ``360probdash: Improving qoe of 360 video streaming using tile-based http adaptive streaming,'' in \emph{Proceedings of the 25th ACM International Conference on Multimedia}.\hskip 1em plus 0.5em minus 0.4em\relax ACM, 2017, pp. 315--323.

\bibitem{lv2020salgcn}
H.~Lv, Q.~Yang, C.~Li, W.~Dai, J.~Zou, and H.~Xiong, ``Salgcn: Saliency prediction for 360-degree images based on spherical graph convolutional networks,'' in \emph{Proceedings of the 28th ACM International Conference on Multimedia}.\hskip 1em plus 0.5em minus 0.4em\relax ACM, 2020, pp. 682--690.

\bibitem{sandler2018mobilenetv2}
M.~Sandler, A.~Howard, M.~Zhu, A.~Zhmoginov, and L.-C. Chen, ``Mobilenetv2: Inverted residuals and linear bottlenecks,'' in \emph{Proceedings of the IEEE conference on computer vision and pattern recognition}, 2018, pp. 4510--4520.

\bibitem{feng2020livedeep}
X.~Feng, Y.~Liu, and S.~Wei, ``Livedeep: Online viewport prediction for live virtual reality streaming using lifelong deep learning,'' in \emph{2020 IEEE Conference on Virtual Reality and 3D User Interfaces (VR)}.\hskip 1em plus 0.5em minus 0.4em\relax IEEE, 2020, pp. 800--808.

\bibitem{he2018rubiks}
J.~He, M.~A. Qureshi, L.~Qiu, J.~Li, F.~Li, and L.~Han, ``Rubiks: Practical 360-degree streaming for smartphones,'' in \emph{Proceedings of the 16th Annual International Conference on Mobile Systems, Applications, and Services}.\hskip 1em plus 0.5em minus 0.4em\relax ACM, 2018, pp. 482--494.

\bibitem{nasrabadi2020viewport}
A.~T. Nasrabadi, A.~Samiei, and R.~Prakash, ``Viewport prediction for 360 videos: a clustering approach,'' in \emph{Proceedings of the 30th ACM Workshop on Network and Operating Systems Support for Digital Audio and Video}, 2020, pp. 34--39.

\bibitem{ban2018cub360}
Y.~Ban, L.~Xie, Z.~Xu, X.~Zhang, Z.~Guo, and Y.~Wang, ``Cub360: Exploiting cross-users behaviors for viewport prediction in 360 video adaptive streaming,'' in \emph{2018 IEEE International Conference on Multimedia and Expo (ICME)}.\hskip 1em plus 0.5em minus 0.4em\relax IEEE, 2018, pp. 1--6.

\bibitem{van2022machine}
S.~Van~Damme, M.~T. Vega, and F.~De~Turck, ``Machine learning based content-agnostic viewport prediction for 360-degree video,'' \emph{ACM Transactions on Multimedia Computing, Communications, and Applications (TOMM)}, vol.~18, no.~2, pp. 1--24, 2022.

\bibitem{xu2019analyzing}
T.~Xu, B.~Han, and F.~Qian, ``Analyzing viewport prediction under different vr interactions,'' in \emph{Proceedings of the 15th International Conference on Emerging Networking Experiments And Technologies}, 2019, pp. 165--171.

\bibitem{fan2017fixation}
C.~Fan, J.~Lee, W.~Lo, C.~Huang, K.~Chen, and C.~Hsu, ``Fixation prediction for 360 video streaming to head-mounted displays,'' \emph{Proceedings of ACM NOSSDAV 2017}, 2017.

\bibitem{xu2018gaze}
Y.~Xu, Y.~Dong, J.~Wu, Z.~Sun, Z.~Shi, J.~Yu, and S.~Gao, ``Gaze prediction in dynamic 360 immersive videos,'' in \emph{proceedings of the IEEE Conference on Computer Vision and Pattern Recognition}, 2018, pp. 5333--5342.

\bibitem{nguyen2018your}
A.~Nguyen, Z.~Yan, and K.~Nahrstedt, ``Your attention is unique: Detecting 360-degree video saliency in head-mounted display for head movement prediction,'' in \emph{Proceedings of the 26th ACM international conference on Multimedia}, 2018, pp. 1190--1198.

\bibitem{li2019very}
C.~Li, W.~Zhang, Y.~Liu, and Y.~Wang, ``Very long term field of view prediction for 360-degree video streaming,'' in \emph{2019 IEEE conference on multimedia information processing and retrieval (MIPR)}.\hskip 1em plus 0.5em minus 0.4em\relax IEEE, 2019, pp. 297--302.

\bibitem{sun2022live}
L.~Sun, Y.~Mao, T.~Zong, Y.~Liu, and Y.~Wang, ``Live 360 degree video delivery based on user collaboration in a streaming flock,'' \emph{IEEE Transactions on Multimedia}, 2022.

\bibitem{bochkovskiy2020yolov4}
A.~Bochkovskiy, C.-Y. Wang, and H.-Y.~M. Liao, ``Yolov4: Optimal speed and accuracy of object detection,'' \emph{arXiv preprint arXiv:2004.10934}, 2020.

\bibitem{wu2020spherical}
C.~Wu, R.~Zhang, Z.~Wang, and L.~Sun, ``A spherical convolution approach for learning long term viewport prediction in 360 immersive video,'' in \emph{Proceedings of the AAAI Conference on Artificial Intelligence}, vol.~34, no.~01, 2020, pp. 14\,003--14\,040.

\bibitem{zhang2019drl360}
Y.~Zhang, P.~Zhao, K.~Bian, Y.~Liu, L.~Song, and X.~Li, ``Drl360: 360-degree video streaming with deep reinforcement learning,'' in \emph{IEEE INFOCOM 2019-IEEE Conference on Computer Communications}.\hskip 1em plus 0.5em minus 0.4em\relax IEEE, 2019, pp. 1252--1260.

\bibitem{chopra2021parima}
L.~Chopra, S.~Chakraborty, A.~Mondal, and S.~Chakraborty, ``Parima: Viewport adaptive 360-degree video streaming,'' in \emph{Proceedings of the Web Conference 2021}, 2021, pp. 2379--2391.

\bibitem{feng2021liveroi}
X.~Feng, W.~Li, and S.~Wei, ``Liveroi: region of interest analysis for viewport prediction in live mobile virtual reality streaming,'' in \emph{Proceedings of the 12th ACM Multimedia Systems Conference}, 2021, pp. 132--145.

\bibitem{feng2021liveobj}
X.~Feng, Z.~Bao, and S.~Wei, ``Liveobj: object semantics-based viewport prediction for live mobile virtual reality streaming,'' \emph{IEEE Transactions on Visualization and Computer Graphics}, vol.~27, no.~5, pp. 2736--2745, 2021.

\bibitem{kummerer2014deep}
M.~K{\"u}mmerer, L.~Theis, and M.~Bethge, ``Deep gaze i: Boosting saliency prediction with feature maps trained on imagenet,'' \emph{arXiv preprint arXiv:1411.1045}, 2014.

\bibitem{pan2016shallow}
J.~Pan, E.~Sayrol, X.~Giro-i Nieto, K.~McGuinness, and N.~E. O'Connor, ``Shallow and deep convolutional networks for saliency prediction,'' in \emph{Proceedings of the IEEE conference on computer vision and pattern recognition}, 2016, pp. 598--606.

\bibitem{zhang2016seed}
Y.~Zhang, L.~Qin, Q.~Huang, K.~Yang, J.~Zhang, and H.~Yao, ``From seed discovery to deep reconstruction: Predicting saliency in crowd via deep networks,'' in \emph{Proceedings of the 24th ACM international conference on Multimedia}, 2016, pp. 72--76.

\bibitem{salomon2006transformations}
D.~Salomon, \emph{Transformations and projections in computer graphics}.\hskip 1em plus 0.5em minus 0.4em\relax Springer, 2006, vol. 233.

\bibitem{yang2020rotation}
Q.~Yang, C.~Li, W.~Dai, J.~Zou, G.-J. Qi, and H.~Xiong, ``Rotation equivariant graph convolutional network for spherical image classification,'' in \emph{Proceedings of the IEEE/CVF Conference on Computer Vision and Pattern Recognition}, 2020, pp. 4303--4312.

\bibitem{ronneberger2015u}
O.~Ronneberger, P.~Fischer, and T.~Brox, ``U-net: Convolutional networks for biomedical image segmentation,'' in \emph{International Conference on Medical image computing and computer-assisted intervention}.\hskip 1em plus 0.5em minus 0.4em\relax Springer, 2015, pp. 234--241.

\bibitem{defferrard2016convolutional}
M.~Defferrard, X.~Bresson, and P.~Vandergheynst, ``Convolutional neural networks on graphs with fast localized spectral filtering,'' \emph{Advances in neural information processing systems}, vol.~29, pp. 3844--3852, 2016.

\bibitem{finn2017model}
C.~Finn, P.~Abbeel, and S.~Levine, ``Model-agnostic meta-learning for fast adaptation of deep networks,'' in \emph{International conference on machine learning}.\hskip 1em plus 0.5em minus 0.4em\relax PMLR, 2017, pp. 1126--1135.

\bibitem{xingjian2015convolutional}
S.~Xingjian, Z.~Chen, H.~Wang, D.-Y. Yeung, W.-K. Wong, and W.-c. Woo, ``Convolutional lstm network: A machine learning approach for precipitation nowcasting,'' in \emph{Advances in neural information processing systems}, 2015, pp. 802--810.

\bibitem{hochreiter1997long}
S.~Hochreiter and J.~Schmidhuber, ``Long short-term memory,'' \emph{Neural computation}, vol.~9, no.~8, pp. 1735--1780, 1997.

\bibitem{wang2018revisiting}
W.~Wang, J.~Shen, F.~Guo, M.-M. Cheng, and A.~Borji, ``Revisiting video saliency: A large-scale benchmark and a new model,'' in \emph{Proceedings of the IEEE Conference on Computer Vision and Pattern Recognition}, 2018, pp. 4894--4903.

\bibitem{guo2017building}
J.~Guo and H.~Chao, ``Building an end-to-end spatial-temporal convolutional network for video super-resolution,'' in \emph{Thirty-First AAAI Conference on Artificial Intelligence}, 2017.

\bibitem{lu2019augur}
Z.~Lu, S.~Rallapalli, K.~Chan, S.~Pu, and T.~La~Porta, ``Augur: Modeling the resource requirements of convnets on mobile devices,'' \emph{IEEE Transactions on Mobile Computing}, vol.~20, no.~2, pp. 352--365, 2019.

\bibitem{szegedy2016rethinking}
C.~Szegedy, V.~Vanhoucke, S.~Ioffe, J.~Shlens, and Z.~Wojna, ``Rethinking the inception architecture for computer vision,'' in \emph{Proceedings of the IEEE conference on computer vision and pattern recognition}, 2016, pp. 2818--2826.

\bibitem{howard2017mobilenets}
A.~G. Howard, M.~Zhu, B.~Chen, D.~Kalenichenko, W.~Wang, T.~Weyand, M.~Andreetto, and H.~Adam, ``Mobilenets: Efficient convolutional neural networks for mobile vision applications,'' \emph{arXiv preprint arXiv:1704.04861}, 2017.

\bibitem{chollet2017xception}
F.~Chollet, ``Xception: Deep learning with depthwise separable convolutions,'' in \emph{Proceedings of the IEEE conference on computer vision and pattern recognition}, 2017, pp. 1251--1258.

\bibitem{hu2018squeeze}
J.~Hu, L.~Shen, and G.~Sun, ``Squeeze-and-excitation networks,'' in \emph{Proceedings of the IEEE conference on computer vision and pattern recognition}, 2018, pp. 7132--7141.

\bibitem{pytorch}
\url{https://pytorch.org/}, Accessed: 2022.

\bibitem{caffe2}
\url{https://caffe2.ai/}, Accessed: 2022.

\bibitem{keras}
\url{https://keras.io/}, Accessed: 2022.

\bibitem{lo2017360}
W.-C. Lo, C.-L. Fan, J.~Lee, C.-Y. Huang, K.-T. Chen, and C.-H. Hsu, ``360 video viewing dataset in head-mounted virtual reality,'' in \emph{Proceedings of the 8th ACM on Multimedia Systems Conference}.\hskip 1em plus 0.5em minus 0.4em\relax ACM, 2017, pp. 211--216.

\bibitem{float16}
``Tensor{F}low,'' \url{https://www.tensorflow.org/lite/performance/post\_training\_float16\_quant}, Accessed: 2022.

\bibitem{van2016http}
J.~Van Der~Hooft, S.~Petrangeli, T.~Wauters, R.~Huysegems, P.~R. Alface, T.~Bostoen, and F.~De~Turck, ``Http/2-based adaptive streaming of hevc video over 4g/lte networks,'' \emph{IEEE Communications Letters}, vol.~20, no.~11, pp. 2177--2180, 2016.

\bibitem{david2018dataset}
E.~J. David, J.~Guti{\'e}rrez, A.~Coutrot, M.~P. Da~Silva, and P.~L. Callet, ``A dataset of head and eye movements for 360\textdegree~videos,'' in \emph{Proceedings of the 9th ACM Multimedia Systems Conference}.\hskip 1em plus 0.5em minus 0.4em\relax ACM, 2018, pp. 432--437.

\bibitem{yao2008empirical}
J.~Yao, S.~S. Kanhere, and M.~Hassan, ``An empirical study of bandwidth predictability in mobile computing,'' in \emph{Proceedings of the third ACM International Workshop on Wireless Network Testbeds, Experimental Evaluation and Characterization}.\hskip 1em plus 0.5em minus 0.4em\relax ACM, 2008, pp. 11--18.

\bibitem{Ma_2018_ECCV}
N.~Ma, X.~Zhang, H.-T. Zheng, and J.~Sun, ``Shufflenet v2: Practical guidelines for efficient cnn architecture design,'' in \emph{Proceedings of the European Conference on Computer Vision (ECCV)}, September 2018.

\bibitem{bouzakaria2014overhead}
N.~Bouzakaria, C.~Concolato, and J.~Le~Feuvre, ``Overhead and performance of low latency live streaming using mpeg-dash,'' in \emph{IISA 2014, The 5th International Conference on Information, Intelligence, Systems and Applications}.\hskip 1em plus 0.5em minus 0.4em\relax IEEE, 2014, pp. 92--97.

\end{thebibliography}
\end{document}